\documentclass[aps,prd,reprint,preprintnumbers,showpacs,nofootinbib,superscript address,longbibliography,floatfix]{revtex4-2}
\usepackage[utf8]{inputenc}
\usepackage{amsmath,amsfonts,amssymb,amsfonts}
\usepackage{fontawesome}
\usepackage{pifont}
\usepackage{mathrsfs}
\usepackage{accents}
\usepackage{verbatim}
\usepackage{natbib}
\usepackage[english]{babel}
\usepackage{adjustbox}
\usepackage{relsize} 
\usepackage{slashed}
\usepackage{scalerel,stackengine}
\usepackage{hhline}
\usepackage{mathtools}
\usepackage[dvipsnames]{xcolor}
\usepackage{longtable}
\usepackage{array}
\usepackage{siunitx}
\usepackage{graphicx}
\usepackage{xstring}
\usepackage{etoolbox}
\usepackage{notoccite}
\usepackage{tensor}
\usepackage{tocbasic}
\DeclareTOCStyleEntry[numwidth=25pt,linefill=\bfseries\TOCLineLeaderFill]{tocline}{section}
\DeclareTOCStyleEntry[entryformat=\textit,numwidth=10pt,linefill=\TOCLineLeaderFill]{tocline}{subsection}
\DeclareTOCStyleEntry[entryformat=\textit,numwidth=10pt,linefill=\TOCLineLeaderFill]{tocline}{subsubsection}
\usepackage{xspace}

\usepackage{tikz}
\usetikzlibrary{calc}
\usepackage{pgfkeys}
\pgfdeclarelayer{foreground}
\pgfsetlayers{main,foreground}

\newrobustcmd{\PSALTer}{\textit{PSALTer}\xspace}
\newrobustcmd{\Wolfram}{\textit{Wolfram}\xspace}

\newrobustcmd{\GenXExpr}[1]{%
	\tensor[^{(#1)}]{X}{}%
}
\newrobustcmd{\GenLExpr}[1]{%
	\tensor[^{(#1)}]{L}{}%
}
\newrobustcmd{\GenNExpr}[1]{%
	\tensor[^{(#1)}]{N}{}%
}

\newrobustcmd{\GenNullVector}[1]{%
	\tensor[^{(#1)}]{\mathsf{v}}{}%
}

\newrobustcmd{\XExpr}[3]{%
	{\tensor[^{({#3})}]{X}{_{{{#1}^{#2}}}}}%
}
\newrobustcmd{\LExpr}[3]{%
	{\tensor[^{({#3})}]{L}{_{{{#1}^{#2}}}}}%
}
\newrobustcmd{\NExpr}[3]{%
	{\tensor[^{({#3})}]{N}{_{{{#1}^{#2}}}}}%
}
\newrobustcmd{\SubXExpr}[4]{%
	{\tensor*[^{({#3})}_{({#4})}]{X}{_{{{#1}^{#2}}}}}%
}
\newrobustcmd{\SubLExpr}[4]{%
	{\tensor*[^{({#3})}_{({#4})}]{L}{_{{{#1}^{#2}}}}}%
}
\newrobustcmd{\SubNExpr}[4]{%
	{\tensor*[^{({#3})}_{({#4})}]{N}{_{{{#1}^{#2}}}}}%
}
\newrobustcmd{\NullVector}[3]{%
	{\tensor[^{(#3)}]{\mathsf{v}}{_{{{#1}^{#2}}}}}%
}

\newrobustcmd{\MAGg}[1]{%
	\tensor{g}{#1}%
}
\newrobustcmd{\G}[1]{%
	\tensor{\eta}{#1}%
}
\newrobustcmd{\MAGF}[1]{%
	\tensor{\mathcal{F}}{#1}%
}
\newrobustcmd{\MAGFFirstThird}[1]{%
	\tensor{\mathcal{F}}{^{(13)}#1}%
}
\newrobustcmd{\MAGA}[1]{%
	\tensor{A}{#1}%
}
\newrobustcmd{\MAGT}[1]{%
	\tensor{\mathcal{T}}{#1}%
}
\newrobustcmd{\Christoffel}[1]{%
	\tensor{\Gamma}{#1}%
}
\newrobustcmd{\Dis}[1]{%
	\tensor{\mathcal{K}}{#1}%
}
\newrobustcmd{\Cur}[1]{%
	{\tensor{\mathcal{J}}{#1}}%
}
\newrobustcmd{\OperatorO}[2]{%
	\tensor*{\mathcal{O}}{^{{#1}}_{(#2)}}
}
\newrobustcmd{\hel}[2][]{%
  \ifstrempty{#1}{%
    \tensor*{\varepsilon}{_{#2}}%
  }{%
    \tensor*{\varepsilon}{_{#2}^{\##1}}%
  }%
}
\newrobustcmd{\MAGCouplingA}[1]{%
	\tensor{a}{_{#1}}%
}
\newrobustcmd{\MAGCouplingG}[1]{%
	\tensor{g}{_{#1}}%
}
\newrobustcmd{\MAGCouplingB}[1]{%
	\tensor{b}{_{#1}}%
}
\newrobustcmd{\MAGG}[2]{%
	\tensor{\vphantom{l^{l^2}}\smash{\stackrel{\raisebox{-3pt}{\scalebox{0.55}{$({#1})$}}}{\scalebox{0.99}{$\kappa$}}}}{#2}%
}
\newrobustcmd{\PD}[1]{%
	\tensor{\partial}{#1}%
}
\newrobustcmd{\LorentzGroup}{%
	\mathrm{SO}^+(3,1)
}

\newrobustcmd{\FinalTally}{22\xspace}
\newrobustcmd{\NamedModel}[1]{\ifstrequal{#1}{A23}{$\aleph$}{\ifstrequal{#1}{A23B1}{$\tensor*{{\mathfrak{B}}}{_{1}}$}{\ifstrequal{#1}{A23B1D1}{$\tensor*{{\mathfrak{D}}}{_{1}}$}{\ifstrequal{#1}{A23B1D1E1}{$\tensor*{{\mathfrak{E}}}{_{14}}$}{\ifstrequal{#1}{A23B1D1E1G1}{$\tensor*{{\mathfrak{G}}}{_{21}}$}{\ifstrequal{#1}{A23B1D1E1G2}{$\tensor*{{\mathfrak{G}}}{_{22}}$}{\ifstrequal{#1}{A23B1D1E1G2H1}{$\tensor*{{\mathfrak{H}}}{_{23}}$}{\ifstrequal{#1}{A23B1D1E1G2H1I1}{$\tensor*{{\mathfrak{I}}}{_{28}}$}{\ifstrequal{#1}{A23B1D1E1G2H1I1K1}{$\tensor*{{\mathfrak{K}}}{_{1}}$}{\ifstrequal{#1}{A23B1D1E1G2H1I1K2}{$\tensor*{{\mathfrak{K}}}{_{2}}$}{\ifstrequal{#1}{A23B1D1E1G2H1J1}{$\tensor*{{\mathfrak{J}}}{_{11}}$}{\ifstrequal{#1}{A23B1D1E1G2H1J1K1}{$\tensor*{{\mathfrak{K}}}{_{3}}$}{\ifstrequal{#1}{A23B1D1E1G2H1J2}{$\tensor*{{\mathfrak{J}}}{_{12}}$}{\ifstrequal{#1}{A23B1D1E1G2I1}{$\tensor*{{\mathfrak{I}}}{_{24}}$}{\ifstrequal{#1}{A23B1D1E1G2I1J1}{$\tensor*{{\mathfrak{J}}}{_{28}}$}{\ifstrequal{#1}{A23B1D1E1G2I2}{$\tensor*{{\mathfrak{I}}}{_{25}}$}{\ifstrequal{#1}{A23B1D1F1}{$\tensor*{{\mathfrak{F}}}{_{4}}$}{\ifstrequal{#1}{A23B1D1F2}{$\tensor*{{\mathfrak{F}}}{_{5}}$}{\ifstrequal{#1}{A23B1D1F2H1}{$\tensor*{{\mathfrak{H}}}{_{7}}$}{\ifstrequal{#1}{A23B1D1F2H1I2}{$\tensor*{{\mathfrak{I}}}{_{33}}$}{\ifstrequal{#1}{A23B1D1F2H2}{$\tensor*{{\mathfrak{H}}}{_{8}}$}{\ifstrequal{#1}{A23B1D2}{$\tensor*{{\mathfrak{D}}}{_{2}}$}{\ifstrequal{#1}{A23B1D2F1}{$\tensor*{{\mathfrak{F}}}{_{6}}$}{\ifstrequal{#1}{A23B1D2F2}{$\tensor*{{\mathfrak{F}}}{_{7}}$}{\ifstrequal{#1}{A23B1D2F2H1}{$\tensor*{{\mathfrak{H}}}{_{9}}$}{\ifstrequal{#1}{A23B1D2F2H1I2}{$\tensor*{{\mathfrak{I}}}{_{34}}$}{\ifstrequal{#1}{A23B1D2F2H2}{$\tensor*{{\mathfrak{H}}}{_{10}}$}{\ifstrequal{#1}{A23B1D3}{$\tensor*{{\mathfrak{D}}}{_{3}}$}{\ifstrequal{#1}{A23B1D3F3}{$\tensor*{{\mathfrak{F}}}{_{8}}$}{\ifstrequal{#1}{A23B1D3F3H1}{$\tensor*{{\mathfrak{H}}}{_{11}}$}{\ifstrequal{#1}{A23B1D3F3H1I1}{$\tensor*{{\mathfrak{I}}}{_{30}}$}{\ifstrequal{#1}{A23B1D3F3H1I1J1}{$\tensor*{{\mathfrak{J}}}{_{27}}$}{\ifstrequal{#1}{A23B1D3F3H1I2}{$\tensor*{{\mathfrak{I}}}{_{29}}$}{\ifstrequal{#1}{A23B1D3F3H2}{$\tensor*{{\mathfrak{H}}}{_{12}}$}{\ifstrequal{#1}{A23B1D3F3H2I2}{$\tensor*{{\mathfrak{I}}}{_{31}}$}{\ifstrequal{#1}{A23B1D3F3H3}{$\tensor*{{\mathfrak{H}}}{_{13}}$}{\ifstrequal{#1}{A23B1D3F3H4}{$\tensor*{{\mathfrak{H}}}{_{14}}$}{\ifstrequal{#1}{A23B1D3F3H4J1}{$\tensor*{{\mathfrak{J}}}{_{1}}$}{\ifstrequal{#1}{A23B1D3F3H4J1K1}{$\tensor*{{\mathfrak{K}}}{_{5}}$}{\ifstrequal{#1}{A23B1D3F3H4J1K2}{$\tensor*{{\mathfrak{K}}}{_{6}}$}{\ifstrequal{#1}{A23B1D3F3H4J1K3}{$\tensor*{{\mathfrak{K}}}{_{7}}$}{\ifstrequal{#1}{A23B1D3F3H4J2}{$\tensor*{{\mathfrak{J}}}{_{2}}$}{\ifstrequal{#1}{A23B1D3F3H4J2K2}{$\tensor*{{\mathfrak{K}}}{_{8}}$}{\ifstrequal{#1}{A23B1D3F3H4J2K3}{$\tensor*{{\mathfrak{K}}}{_{9}}$}{\ifstrequal{#1}{A23B1D3F3H4J3}{$\tensor*{{\mathfrak{J}}}{_{3}}$}{\ifstrequal{#1}{A23B1D3F3H4J3K3}{$\tensor*{{\mathfrak{K}}}{_{10}}$}{\ifstrequal{#1}{A23B1D3F3H4J4}{$\tensor*{{\mathfrak{J}}}{_{4}}$}{\ifstrequal{#1}{A23B1D3F4}{$\tensor*{{\mathfrak{F}}}{_{9}}$}{\ifstrequal{#1}{A23B1D3F4H1}{$\tensor*{{\mathfrak{H}}}{_{15}}$}{\ifstrequal{#1}{A23B1D3F4H1I2}{$\tensor*{{\mathfrak{I}}}{_{35}}$}{\ifstrequal{#1}{A23B1D3F4H2}{$\tensor*{{\mathfrak{H}}}{_{16}}$}{\ifstrequal{#1}{A23B1D3F4H3}{$\tensor*{{\mathfrak{H}}}{_{17}}$}{\ifstrequal{#1}{A23B1D4}{$\tensor*{{\mathfrak{D}}}{_{4}}$}{\ifstrequal{#1}{A23B1D4F1}{$\tensor*{{\mathfrak{F}}}{_{10}}$}{\ifstrequal{#1}{A23B1D4F1G1}{$\tensor*{{\mathfrak{G}}}{_{28}}$}{\ifstrequal{#1}{A23B1D4F1G1H1}{$\tensor*{{\mathfrak{H}}}{_{34}}$}{\ifstrequal{#1}{A23B1D4F1G2}{$\tensor*{{\mathfrak{G}}}{_{29}}$}{\ifstrequal{#1}{A23B1D4F1H2}{$\tensor*{{\mathfrak{H}}}{_{18}}$}{\ifstrequal{#1}{A23B1D4F1H2J1}{$\tensor*{{\mathfrak{J}}}{_{5}}$}{\ifstrequal{#1}{A23B1D4F1H2J2}{$\tensor*{{\mathfrak{J}}}{_{6}}$}{\ifstrequal{#1}{A23B1D4F2}{$\tensor*{{\mathfrak{F}}}{_{11}}$}{\ifstrequal{#1}{A23B1D4F2G2}{$\tensor*{{\mathfrak{G}}}{_{30}}$}{\ifstrequal{#1}{A23B1D4F2H1}{$\tensor*{{\mathfrak{H}}}{_{19}}$}{\ifstrequal{#1}{A23B1D4F2H1J1}{$\tensor*{{\mathfrak{J}}}{_{7}}$}{\ifstrequal{#1}{A23B1D4F2H1J2}{$\tensor*{{\mathfrak{J}}}{_{8}}$}{\ifstrequal{#1}{A23B1D4F3}{$\tensor*{{\mathfrak{F}}}{_{12}}$}{\ifstrequal{#1}{A23B1D4F3H2}{$\tensor*{{\mathfrak{H}}}{_{20}}$}{\ifstrequal{#1}{A23B1D4F3H2J1}{$\tensor*{{\mathfrak{J}}}{_{9}}$}{\ifstrequal{#1}{A23B1D4F3H2J2}{$\tensor*{{\mathfrak{J}}}{_{10}}$}{\ifstrequal{#1}{A23B1D4G1}{$\tensor*{{\mathfrak{G}}}{_{1}}$}{\ifstrequal{#1}{A23B1D4G1I1}{$\tensor*{{\mathfrak{I}}}{_{1}}$}{\ifstrequal{#1}{A23B1D4G1I2}{$\tensor*{{\mathfrak{I}}}{_{2}}$}{\ifstrequal{#1}{A23B1D4G2}{$\tensor*{{\mathfrak{G}}}{_{2}}$}{\ifstrequal{#1}{A23B1D4G2I1}{$\tensor*{{\mathfrak{I}}}{_{3}}$}{\ifstrequal{#1}{A23B1D4G2I1J2}{$\tensor*{{\mathfrak{J}}}{_{21}}$}{\ifstrequal{#1}{A23B1D4G2I1J3}{$\tensor*{{\mathfrak{J}}}{_{22}}$}{\ifstrequal{#1}{A23B1D4G2I3}{$\tensor*{{\mathfrak{I}}}{_{4}}$}{\ifstrequal{#1}{A23B1D4G2I3J3}{$\tensor*{{\mathfrak{J}}}{_{23}}$}{\ifstrequal{#1}{A23B1D4G2I4}{$\tensor*{{\mathfrak{I}}}{_{5}}$}{\ifstrequal{#1}{A23B1E1}{$\tensor*{{\mathfrak{E}}}{_{1}}$}{\ifstrequal{#1}{A23B1E1G1}{$\tensor*{{\mathfrak{G}}}{_{5}}$}{\ifstrequal{#1}{A23B1E1G2}{$\tensor*{{\mathfrak{G}}}{_{6}}$}{\ifstrequal{#1}{A23B1E2}{$\tensor*{{\mathfrak{E}}}{_{2}}$}{\ifstrequal{#1}{A23B1E2G1}{$\tensor*{{\mathfrak{G}}}{_{7}}$}{\ifstrequal{#1}{A23B1E2G1H2}{$\tensor*{{\mathfrak{H}}}{_{33}}$}{\ifstrequal{#1}{A23B1E2G4}{$\tensor*{{\mathfrak{G}}}{_{8}}$}{\ifstrequal{#1}{A23B2}{$\tensor*{{\mathfrak{B}}}{_{2}}$}{\ifstrequal{#1}{A23B2C1}{$\tensor*{{\mathfrak{C}}}{_{5}}$}{\ifstrequal{#1}{A23B2C1E2}{$\tensor*{{\mathfrak{E}}}{_{15}}$}{\ifstrequal{#1}{A23B2C1E2G2}{$\tensor*{{\mathfrak{G}}}{_{23}}$}{\ifstrequal{#1}{A23B2C1E2G2H1}{$\tensor*{{\mathfrak{H}}}{_{26}}$}{\ifstrequal{#1}{A23B2C1E2G2H1I1}{$\tensor*{{\mathfrak{I}}}{_{32}}$}{\ifstrequal{#1}{A23B2C1E2G2H1I1K2}{$\tensor*{{\mathfrak{K}}}{_{4}}$}{\ifstrequal{#1}{A23B2C1E2G2H1J2}{$\tensor*{{\mathfrak{J}}}{_{14}}$}{\ifstrequal{#1}{A23B2C1E2G2I1}{$\tensor*{{\mathfrak{I}}}{_{26}}$}{\ifstrequal{#1}{A23B2C1E2G2I2}{$\tensor*{{\mathfrak{I}}}{_{27}}$}{\ifstrequal{#1}{A23B2C1E3}{$\tensor*{{\mathfrak{E}}}{_{16}}$}{\ifstrequal{#1}{A23B2C1E3F1}{$\tensor*{{\mathfrak{F}}}{_{16}}$}{\ifstrequal{#1}{A23B2C1E3F1H2}{$\tensor*{{\mathfrak{H}}}{_{25}}$}{\ifstrequal{#1}{A23B2C1E3F1H2J2}{$\tensor*{{\mathfrak{J}}}{_{13}}$}{\ifstrequal{#1}{A23B2C1E3F1H4}{$\tensor*{{\mathfrak{H}}}{_{24}}$}{\ifstrequal{#1}{A23B2C1E3G1}{$\tensor*{{\mathfrak{G}}}{_{24}}$}{\ifstrequal{#1}{A23B2C1E3G3}{$\tensor*{{\mathfrak{G}}}{_{25}}$}{\ifstrequal{#1}{A23B2C1E4}{$\tensor*{{\mathfrak{E}}}{_{17}}$}{\ifstrequal{#1}{A23B2C1E4G2}{$\tensor*{{\mathfrak{G}}}{_{26}}$}{\ifstrequal{#1}{A23B2C1E4G2H1}{$\tensor*{{\mathfrak{H}}}{_{35}}$}{\ifstrequal{#1}{A23B2C1E4G3}{$\tensor*{{\mathfrak{G}}}{_{27}}$}{\ifstrequal{#1}{A23B2D2}{$\tensor*{{\mathfrak{D}}}{_{5}}$}{\ifstrequal{#1}{A23B2D2F2}{$\tensor*{{\mathfrak{F}}}{_{13}}$}{\ifstrequal{#1}{A23B2D2F2H1}{$\tensor*{{\mathfrak{H}}}{_{21}}$}{\ifstrequal{#1}{A23B2D2F2H2}{$\tensor*{{\mathfrak{H}}}{_{22}}$}{\ifstrequal{#1}{A23B2D3}{$\tensor*{{\mathfrak{D}}}{_{6}}$}{\ifstrequal{#1}{A23B2D3F1}{$\tensor*{{\mathfrak{F}}}{_{14}}$}{\ifstrequal{#1}{A23B2D3F2}{$\tensor*{{\mathfrak{F}}}{_{15}}$}{\ifstrequal{#1}{A23B2D4}{$\tensor*{{\mathfrak{D}}}{_{7}}$}{\ifstrequal{#1}{A23B2D4F2}{$\tensor*{{\mathfrak{F}}}{_{17}}$}{\ifstrequal{#1}{A23B2D4F2G2}{$\tensor*{{\mathfrak{G}}}{_{31}}$}{\ifstrequal{#1}{A23B2D4F4}{$\tensor*{{\mathfrak{F}}}{_{18}}$}{\ifstrequal{#1}{A23B3}{$\tensor*{{\mathfrak{B}}}{_{3}}$}{\ifstrequal{#1}{A23B3D2}{$\tensor*{{\mathfrak{D}}}{_{8}}$}{\ifstrequal{#1}{A23B3D2F2}{$\tensor*{{\mathfrak{F}}}{_{19}}$}{\ifstrequal{#1}{A23B3D2F2H1}{$\tensor*{{\mathfrak{H}}}{_{27}}$}{\ifstrequal{#1}{A23B3D2F2H2}{$\tensor*{{\mathfrak{H}}}{_{28}}$}{\ifstrequal{#1}{A23B3D3}{$\tensor*{{\mathfrak{D}}}{_{9}}$}{\ifstrequal{#1}{A23B3D3F1}{$\tensor*{{\mathfrak{F}}}{_{20}}$}{\ifstrequal{#1}{A23B3D3F4}{$\tensor*{{\mathfrak{F}}}{_{21}}$}{\ifstrequal{#1}{A23B3D4}{$\tensor*{{\mathfrak{D}}}{_{10}}$}{\ifstrequal{#1}{A23B3D4F3}{$\tensor*{{\mathfrak{F}}}{_{22}}$}{\ifstrequal{#1}{A23B3D4F3G2}{$\tensor*{{\mathfrak{G}}}{_{32}}$}{\ifstrequal{#1}{A23B3D4F4}{$\tensor*{{\mathfrak{F}}}{_{23}}$}{\ifstrequal{#1}{A23B4}{$\tensor*{{\mathfrak{B}}}{_{4}}$}{\ifstrequal{#1}{A23B4D4}{$\tensor*{{\mathfrak{D}}}{_{11}}$}{\ifstrequal{#1}{A23B4D4F2}{$\tensor*{{\mathfrak{F}}}{_{24}}$}{\ifstrequal{#1}{A23B4D4F2G2}{$\tensor*{{\mathfrak{G}}}{_{33}}$}{\ifstrequal{#1}{A23B4D4F2H2}{$\tensor*{{\mathfrak{H}}}{_{29}}$}{\ifstrequal{#1}{A23B4D4F2H2J1}{$\tensor*{{\mathfrak{J}}}{_{15}}$}{\ifstrequal{#1}{A23B4D4F2H2J2}{$\tensor*{{\mathfrak{J}}}{_{16}}$}{\ifstrequal{#1}{A23B4D4F3}{$\tensor*{{\mathfrak{F}}}{_{25}}$}{\ifstrequal{#1}{A23B4D4F3H2}{$\tensor*{{\mathfrak{H}}}{_{30}}$}{\ifstrequal{#1}{A23B4D4F3H2J1}{$\tensor*{{\mathfrak{J}}}{_{17}}$}{\ifstrequal{#1}{A23B4D4F3H2J2}{$\tensor*{{\mathfrak{J}}}{_{18}}$}{\ifstrequal{#1}{A23B4D4F4}{$\tensor*{{\mathfrak{F}}}{_{26}}$}{\ifstrequal{#1}{A23B4D4F4H2}{$\tensor*{{\mathfrak{H}}}{_{31}}$}{\ifstrequal{#1}{A23B4D4F4H2J1}{$\tensor*{{\mathfrak{J}}}{_{19}}$}{\ifstrequal{#1}{A23B4D4F4H2J2}{$\tensor*{{\mathfrak{J}}}{_{20}}$}{\ifstrequal{#1}{A23B4D4G1}{$\tensor*{{\mathfrak{G}}}{_{3}}$}{\ifstrequal{#1}{A23B4D4G1I1}{$\tensor*{{\mathfrak{I}}}{_{6}}$}{\ifstrequal{#1}{A23B4D4G1I2}{$\tensor*{{\mathfrak{I}}}{_{7}}$}{\ifstrequal{#1}{A23B4D4G2}{$\tensor*{{\mathfrak{G}}}{_{4}}$}{\ifstrequal{#1}{A23B4D4G2I1}{$\tensor*{{\mathfrak{I}}}{_{8}}$}{\ifstrequal{#1}{A23B4D4G2I1J2}{$\tensor*{{\mathfrak{J}}}{_{24}}$}{\ifstrequal{#1}{A23B4D4G2I1J4}{$\tensor*{{\mathfrak{J}}}{_{25}}$}{\ifstrequal{#1}{A23B4D4G2I2}{$\tensor*{{\mathfrak{I}}}{_{9}}$}{\ifstrequal{#1}{A23B4D4G2I2J4}{$\tensor*{{\mathfrak{J}}}{_{26}}$}{\ifstrequal{#1}{A23B4D4G2I3}{$\tensor*{{\mathfrak{I}}}{_{10}}$}{\ifstrequal{#1}{A23B4E1}{$\tensor*{{\mathfrak{E}}}{_{3}}$}{\ifstrequal{#1}{A23B4E1G1}{$\tensor*{{\mathfrak{G}}}{_{9}}$}{\ifstrequal{#1}{A23B4E1G2}{$\tensor*{{\mathfrak{G}}}{_{10}}$}{\ifstrequal{#1}{A23B4E2}{$\tensor*{{\mathfrak{E}}}{_{4}}$}{\ifstrequal{#1}{A23B4E2G1}{$\tensor*{{\mathfrak{G}}}{_{11}}$}{\ifstrequal{#1}{A23B4E2G1H2}{$\tensor*{{\mathfrak{H}}}{_{32}}$}{\ifstrequal{#1}{A23B4E2G2}{$\tensor*{{\mathfrak{G}}}{_{12}}$}{\ifstrequal{#1}{A23C1}{$\tensor*{{\mathfrak{C}}}{_{1}}$}{\ifstrequal{#1}{A23C1E1}{$\tensor*{{\mathfrak{E}}}{_{5}}$}{\ifstrequal{#1}{A23C1E1G1}{$\tensor*{{\mathfrak{G}}}{_{13}}$}{\ifstrequal{#1}{A23C1E1G1I1}{$\tensor*{{\mathfrak{I}}}{_{11}}$}{\ifstrequal{#1}{A23C1E1G1I2}{$\tensor*{{\mathfrak{I}}}{_{12}}$}{\ifstrequal{#1}{A23C1E2}{$\tensor*{{\mathfrak{E}}}{_{6}}$}{\ifstrequal{#1}{A23C1E2G1}{$\tensor*{{\mathfrak{G}}}{_{14}}$}{\ifstrequal{#1}{A23C1E2G1I1}{$\tensor*{{\mathfrak{I}}}{_{13}}$}{\ifstrequal{#1}{A23C1E2G1I2}{$\tensor*{{\mathfrak{I}}}{_{14}}$}{\ifstrequal{#1}{A23C1E3}{$\tensor*{{\mathfrak{E}}}{_{7}}$}{\ifstrequal{#1}{A23C1E3G2}{$\tensor*{{\mathfrak{G}}}{_{15}}$}{\ifstrequal{#1}{A23C1E3G3}{$\tensor*{{\mathfrak{G}}}{_{16}}$}{\ifstrequal{#1}{A23C2}{$\tensor*{{\mathfrak{C}}}{_{2}}$}{\ifstrequal{#1}{A23C2E1}{$\tensor*{{\mathfrak{E}}}{_{8}}$}{\ifstrequal{#1}{A23C2E1F3}{$\tensor*{{\mathfrak{F}}}{_{27}}$}{\ifstrequal{#1}{A23C2E1G2}{$\tensor*{{\mathfrak{G}}}{_{17}}$}{\ifstrequal{#1}{A23C2E1G2I1}{$\tensor*{{\mathfrak{I}}}{_{15}}$}{\ifstrequal{#1}{A23C2E1G2I2}{$\tensor*{{\mathfrak{I}}}{_{16}}$}{\ifstrequal{#1}{A23C2E3}{$\tensor*{{\mathfrak{E}}}{_{9}}$}{\ifstrequal{#1}{A23C2E3G2}{$\tensor*{{\mathfrak{G}}}{_{18}}$}{\ifstrequal{#1}{A23C2E3G2I1}{$\tensor*{{\mathfrak{I}}}{_{17}}$}{\ifstrequal{#1}{A23C2E3G2I2}{$\tensor*{{\mathfrak{I}}}{_{18}}$}{\ifstrequal{#1}{A23C2F1}{$\tensor*{{\mathfrak{F}}}{_{1}}$}{\ifstrequal{#1}{A23C2F1H1}{$\tensor*{{\mathfrak{H}}}{_{1}}$}{\ifstrequal{#1}{A23C2F1H1I3}{$\tensor*{{\mathfrak{I}}}{_{20}}$}{\ifstrequal{#1}{A23C2F1H1I4}{$\tensor*{{\mathfrak{I}}}{_{23}}$}{\ifstrequal{#1}{A23C2F1H2}{$\tensor*{{\mathfrak{H}}}{_{2}}$}{\ifstrequal{#1}{A23C2F1H3}{$\tensor*{{\mathfrak{H}}}{_{3}}$}{\ifstrequal{#1}{A23C2F1H3I4}{$\tensor*{{\mathfrak{I}}}{_{22}}$}{\ifstrequal{#1}{A23C2F1H4}{$\tensor*{{\mathfrak{H}}}{_{4}}$}{\ifstrequal{#1}{A23C2F2}{$\tensor*{{\mathfrak{F}}}{_{2}}$}{\ifstrequal{#1}{A23C2F2H2}{$\tensor*{{\mathfrak{H}}}{_{5}}$}{\ifstrequal{#1}{A23C3}{$\tensor*{{\mathfrak{C}}}{_{3}}$}{\ifstrequal{#1}{A23C3E2}{$\tensor*{{\mathfrak{E}}}{_{10}}$}{\ifstrequal{#1}{A23C3E4}{$\tensor*{{\mathfrak{E}}}{_{11}}$}{\ifstrequal{#1}{A23C4}{$\tensor*{{\mathfrak{C}}}{_{4}}$}{\ifstrequal{#1}{A23C4E2}{$\tensor*{{\mathfrak{E}}}{_{12}}$}{\ifstrequal{#1}{A23C4E2F2}{$\tensor*{{\mathfrak{F}}}{_{28}}$}{\ifstrequal{#1}{A23C4E2G2}{$\tensor*{{\mathfrak{G}}}{_{19}}$}{\ifstrequal{#1}{A23C4E2G2I1}{$\tensor*{{\mathfrak{I}}}{_{19}}$}{\ifstrequal{#1}{A23C4E4}{$\tensor*{{\mathfrak{E}}}{_{13}}$}{\ifstrequal{#1}{A23C4E4G2}{$\tensor*{{\mathfrak{G}}}{_{20}}$}{\ifstrequal{#1}{A23C4E4G2I1}{$\tensor*{{\mathfrak{I}}}{_{21}}$}{\ifstrequal{#1}{A23C4F2}{$\tensor*{{\mathfrak{F}}}{_{3}}$}{\ifstrequal{#1}{A23C4F2H1}{$\tensor*{{\mathfrak{H}}}{_{6}}$}{\ifstrequal{#1}{A23Z1}{$\varnothing$}{$\tensor*{{\mathfrak{X}}}{_{0}}$}}}}}}}}}}}}}}}}}}}}}}}}}}}}}}}}}}}}}}}}}}}}}}}}}}}}}}}}}}}}}}}}}}}}}}}}}}}}}}}}}}}}}}}}}}}}}}}}}}}}}}}}}}}}}}}}}}}}}}}}}}}}}}}}}}}}}}}}}}}}}}}}}}}}}}}}}}}}}}}}}}}}}}}}}}}}}}}}}}}}}}}}}}}}}}}}}}}}}}}}}}}}}}}}}
\newrobustcmd{\Lagrangian}[1]{%
	{\tensor*{\mathcal{L}}{_{\text{\NamedModel{#1}}}}}
}
\newrobustcmd{\UnknownMasslessParticle}{%
	{\big\{\tensor*{J}{^{P}_{\gamma}}\big\}}%
}
\newrobustcmd{\MasslessParticle}[2]{%
	{\tensor*{#1}{^{#2}_{\gamma}}}
}
\newrobustcmd{\MassiveParticle}[2]{%
	{\tensor*{#1}{^{#2}_{\text{m}}}}
}

\pgfkeys{
  /AnnotatedGraph/.is family, /AnnotatedGraph,
  A/.code n args={3}{ \node (labelA) at (#1, #3) {\huge$\mathfrak{A}_i$}; \draw[->,thick] (labelA.south) -- (#1, #2); },
  B/.code n args={3}{ \node (labelB) at (#1, #3) {\huge$\mathfrak{B}_i$}; \draw[->,thick] (labelB.south) -- (#1, #2); },
  C/.code n args={3}{ \node (labelC) at (#1, #3) {\huge$\mathfrak{C}_i$}; \draw[->,thick] (labelC.south) -- (#1, #2); },
  D/.code n args={3}{ \node (labelD) at (#1, #3) {\huge$\mathfrak{D}_i$}; \draw[->,thick] (labelD.south) -- (#1, #2); },
  E/.code n args={3}{ \node (labelE) at (#1, #3) {\huge$\mathfrak{E}_i$}; \draw[->,thick] (labelE.south) -- (#1, #2); },
  F/.code n args={3}{ \node (labelF) at (#1, #3) {\huge$\mathfrak{F}_i$}; \draw[->,thick] (labelF.south) -- (#1, #2); },
  G/.code n args={3}{ \node (labelG) at (#1, #3) {\huge$\mathfrak{G}_i$}; \draw[->,thick] (labelG.south) -- (#1, #2); },
  H/.code n args={3}{ \node (labelH) at (#1, #3) {\huge$\mathfrak{H}_i$}; \draw[->,thick] (labelH.south) -- (#1, #2); },
  I/.code n args={3}{ \node (labelI) at (#1, #3) {\huge$\mathfrak{I}_i$}; \draw[->,thick] (labelI.south) -- (#1, #2); },
  J/.code n args={3}{ \node (labelJ) at (#1, #3) {\huge$\mathfrak{J}_i$}; \draw[->,thick] (labelJ.south) -- (#1, #2); },
  K/.code n args={3}{ \node (labelK) at (#1, #3) {\huge$\mathfrak{K}_i$}; \draw[->,thick] (labelK.south) -- (#1, #2); },
  L/.code n args={3}{ \node (labelL) at (#1, #3) {\huge$\mathfrak{L}_i$}; \draw[->,thick] (labelL.south) -- (#1, #2); },
  M/.code n args={3}{ \node (labelM) at (#1, #3) {\huge$\mathfrak{M}_i$}; \draw[->,thick] (labelM.south) -- (#1, #2); },
  N/.code n args={3}{ \node (labelN) at (#1, #3) {\huge$\mathfrak{N}_i$}; \draw[->,thick] (labelN.south) -- (#1, #2); },
}

\newrobustcmd{\AnnotatedGraph}[2][]{%
  \begin{tikzpicture}
    \node[anchor=south west, inner sep=0] (image) at (0,0) {\includegraphics[width=\linewidth]{#2}};

    \begin{pgfonlayer}{foreground}
        \begin{scope}[
            x={($(image.south east)-(image.south west)$)},
            y={($(image.north west)-(image.south west)$)}
        ]
          \pgfkeys{/AnnotatedGraph, #1}
        \end{scope}
    \end{pgfonlayer}
  \end{tikzpicture}%
}

\usepackage{hyperref}
\hypersetup{%
colorlinks = true,%
linkcolor = Red,%
citecolor = Red,%
filecolor = Red,%
urlcolor = Red%
}%
\usepackage[capitalize]{cleveref}
\AddToHook{cmd/appendix/before}{\crefalias{section}{appendix}}
\makeatletter
\renewcommand{\paragraph}{%
\@startsection{paragraph}{4}%
{\z@}{1.21ex \@plus 1ex \@minus .2ex}{0.9em}%
{\normalfont\normalsize\bfseries}%
}
\usepackage{titlesec}
\newrobustcmd{\pea}[1]{\emph{#1}\textbf{.\ \ \ ---}}
\titleformat{\paragraph}[runin]{\normalfont\normalsize\bfseries}{\emph\theparagraph}{1em}{\pea}
\titleformat{\section}[block]{\normalfont\bfseries\centering}{\MakeUppercase\thesection}{1em}{\MakeUppercase}
\makeatother

\makeatletter \def\switch@array{}\makeatother

\allowdisplaybreaks

\begin{document}

\title{Infrared foundations for quantum geometry II:\\
Catalogue of all torsion-like theories including new ghost-tachyon-free cases}

\author{Will Barker}
\email{barker@fzu.cz}
\affiliation{Central European Institute for Cosmology and Fundamental Physics, Institute of Physics of the Czech Academy of Sciences, Na Slovance 1999/2, 182 00 Prague 8, Czechia}

\author{Carlo Marzo}
\email{carlo.marzo@kbfi.ee}
\affiliation{Laboratory for High Energy and Computational Physics, NICPB, R\"{a}vala 10, Tallinn 10143, Estonia}

\author{Alessandro Santoni}
\email{asantoni@uc.cl}
\affiliation{Institut f\"ur Theoretische Physik, Technische Universit\"at Wien, Wiedner Hauptstrasse 8--10, A-1040 Vienna, Austria}
\affiliation{Facultad de F\'isica, Pontificia Universidad Cat\'olica de Chile, Vicu\~{n}a Mackenna 4860, Santiago, Chile}

\begin{abstract}
	The construction of consistent effective field theories in the infrared demands that models be defined by their underlying gauge symmetries, rather than by an arbitrary tuning of couplings or a cherry-picking of operators which may not be stable against radiative corrections. Adhering to this principle, we systematically derive all linear, parity-conserving models that propagate a pair-antisymmetric rank-three field on a Minkowski background. Such models are relevant not only to torsion, but to many areas in high-energy physics ranging from dual graviton formulations to string theory and higher-spin theories. Following this exhaustive classification, we extract several unitary models. In the context of torsion, the results are remarkable. None of the models we obtain propagate scalar or pseudoscalar torsion, in stark contrast to the literature focus. Instead, all models propagate one or more vector torsion modes.
\end{abstract}

\maketitle

\tableofcontents

\begin{figure}[h]
\includegraphics[width=\linewidth]{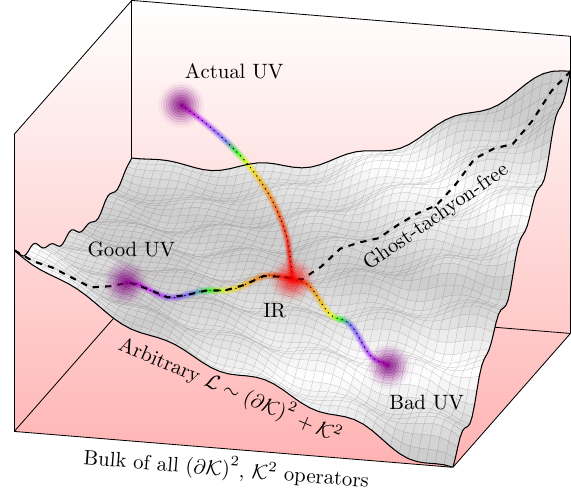}
\caption{\label{Propaganda} The stability of a low-energy theory against radiative corrections is not guaranteed if its operator content is chosen arbitrarily. Any fine-tuning of couplings to ensure a ghost- and tachyon-free spectrum is generally undone by quantum loops. These issues are avoided when the model is defined by a gauge symmetry, which protects the physical properties of the theory in a robust and principled manner.}
\end{figure}

\section{Introduction}\label{Sec:Introduction}

\begin{figure}[htbp]
	\includegraphics[width=\linewidth]{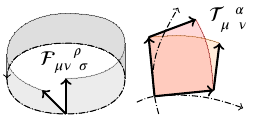}
	\caption{The pair-antisymmetric rank-three field~$\Dis{^\alpha_{\beta\chi}}$ in~\cref{eq:DisA23} has a natural interpretation through~\cref{MAGTDef,postriem,Contorsion} as the spacetime torsion tensor~$\MAGT{_{\mu}^\alpha_{\nu}}$, which encodes an alternative geometric property of spacetime (non-closure of infinitesimally-parallel-transported vectors) to the usual curvature~$\MAGF{_{\mu\nu}^\rho_\sigma}$ in~\cref{MAGFDef} (rotation of a vector upon parallel transport around a loop). As shown in~\cite{Barker:2025xzd}, the universal infrared limit of propagating torsion is~\cref{RootTheoryA23} in which `true' gravity (the metric perturbation) pays no role whatsoever.}
\label{NonRiemannianSchematic}
\end{figure}

\paragraph*{Consistent quantum theories} The principles of effective field theory (EFT), as detailed in~\cite{Barker:2025xzd,Barker:2025rzd} and illustrated in~\cref{Propaganda}, dictate that a viable model for a field~$\Dis{}$ must be constructed from the complete set of operators compatible with its underlying symmetries. This construction is carried out via a perturbative approach, which is fundamentally a low-energy expansion organised in powers of the canonically normalised~$\Dis{}$ and its derivatives
\begin{align}\label{EFTLag}
\mathcal{L}=\mathcal{L}_{n\leq4}+\sum_{n=5}^{\infty}\sum_i\frac{\MAGG{n}{_i}\OperatorO{i}{n}}{\Lambda^{n-4}} \,.
\end{align}
In this expansion, the coefficients~$\big\{\MAGG{n}{_i}\big\}$ are dimensionless couplings, where~$i$ indexes operators of the same mass dimension~$n$, and~$\Lambda$ serves as the theory's cutoff scale. Assuming the field~$\Dis{}$ has a canonical mass dimension of one and that no higher derivatives are present, the bootstrap analysis starts with the Lagrangian~$\mathcal{L}_{n\leq4}$ in~\cref{EFTLag}. This term includes all relevant and marginal operators encompassing possible cubic and quartic interactions, as well as the dominant part, responsible for propagation, quadratic in~$\Dis{}$. For a high-rank tensor field, this general form of~$\mathcal{L}_{n\leq4}$ is typically non-unitary in the absence of an additional gauge symmetry. This underscores a critical point: a self-consistent theory cannot be built upon the field~$\Dis{}$ alone. Instead, a foundational gauge principle is necessary, from which a unitary model can be uniquely derived. This `symmetrization' approach must be contrasted with the ad hoc `cherry-picking' of operators or fine-tuning of couplings to manually achieve a healthy spectrum of propagating degrees of freedom (d.o.f). Such arbitrary adjustments are not generally stable under radiative corrections and risk yielding a model that is not even systematically renormalizable, thus lacking any predictive power within the framework of~\cref{EFTLag}~\cite{Georgi:1993mps,Burgess:2007pt,Bijnens:2006zp,Manohar:1996cq,Kaplan:1995uv,Pich:1995bw,Ecker:1994gg,Dobado:1989ax,Dobado:1989gr,Weinberg:1978kz,Coleman:1969sm,Callan:1969sn}.

\begin{table*}[htbp]
	\includegraphics[width=\linewidth]{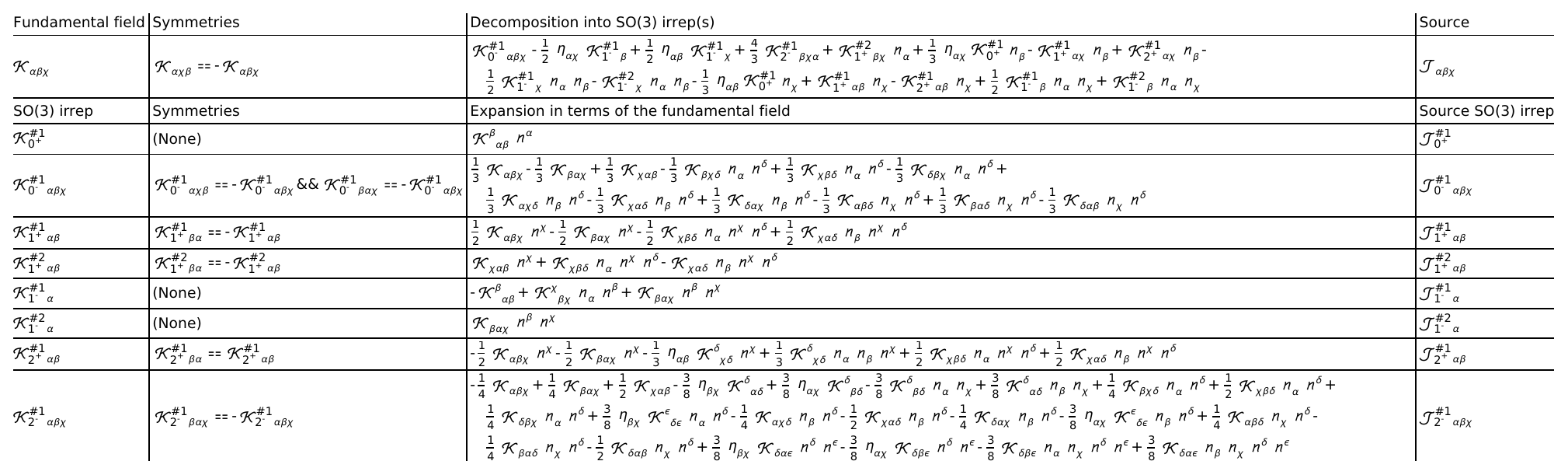}
	\caption{Output generated by \PSALTer{}. Kinematic decomposition of the pair-antisymmetric field from~\cref{eq:DisA23}, providing the basis for the spectrographs in~\cref{ParticleSpectrographA23,ParticleSpectrographA23B1D1E1G2H1I1K2,ParticleSpectrographA23B1D1E1G2H1J1K1,ParticleSpectrographA23B1D1E1G2H1J2}. The unit-timelike vector~$\tensor{n}{_{\mu}}\equiv\tensor{k}{_{\mu}}/\sqrt{\tensor{k}{^\nu}\tensor{k}{_\nu}}$ is constructed using the four-momentum~$\tensor{k}{_\mu}$ of a massive particle.}
	\label{FieldKinematicsA23Field}
\end{table*}

\paragraph*{Pair-antisymmetric rank-three} Our particular choice for~$\Dis{}$ is the pair-antisymmetric rank-three field of the form
\begin{equation}\label{eq:DisA23}
	\Dis{_{\alpha\beta\gamma}}\equiv- \Dis{_{\alpha\gamma\beta}} \, .
\end{equation}
The high-energy physics literature provides abundant motivations for model-building from~\cref{eq:DisA23}. Such a field may be endowed with a variety of interrelated physical interpretations (see the review~\cite{Danehkar2018}):
\begin{description}
	\item[Dual graviton] As a Curtright field, i.e. a gauge potential whose linearized field strength is dual to the Fierz--Pauli graviton curvature~\cite{Curtright1980}. In higher dimensions, this field carries the same on-shell d.o.f as a massless spin-two field, but in four dimensions its Weyl-like curvature vanishes identically~\cite{Deser1981,Hull2001}. A massive extension was proposed in~\cite{CurtrightFreund1981}.\footnote{Properly, the Curtright field should additionally specify~\cref{eq:DisA23} to have no totally antisymmetric part.}
\item[Monopoles and NUT charge] As the natural potential conjugate to Taub--NUT (magnetic) mass; in analogy to how a magnetic monopole sources a dual vector potential~\cite{Hull2024}. The field is sourced by the magnetic energy-momentum tensor, i.e. the stress-energy carried by the Dirac string or gravitational NUT charge, furnishing a local action principle and charge-quantization condition for gravitomagnetic charge~\cite{BunsterHenneaux2012,BunsterEtAl2006}.
\item[Lanczos potential] As the Lanczos tensor, which serves as a potential for the Weyl tensor, providing a first-order, potential-based description of the free gravitational field in four dimensions~\cite{Lanczos1962}.
\item[String theory and higher spins] As one of an infinite tower of fields appearing in the tensionless limit of string theory (and in high-level excitations). The pair-antisymmetric case may arise along with fields of mixed symmetry~\cite{Sagnotti2004,Campoleoni:2008jq}. In higher-spin theory the pair-antisymmetric field also plays a role as a gauge generator~\cite{BekaertBoulanger2003}.
\item[Unified theories] As the dual graviton of mixed symmetry which would arise alongside the four-dimensional metric in West's~$\mathrm{E}_{11}$ Kac--Moody approach to M-theory. Such a field is required by extended symmetry algebras~\cite{West2001}.
\end{description}
Following the earlier parts of this series in~\cite{Barker:2025xzd,Barker:2025rzd}, in this letter we will focus, not on the above applications of~\cref{eq:DisA23}, but (somewhat arbitrarily) on the application to spacetime torsion. Despite this focus, we emphasise that the actual science product, the catalogue of models derived below, is relevant to a much broader audience.

\paragraph*{Application to torsion} Compared to the totally symmetric rank-three field considered in~\cite{Barker:2025xzd,Barker:2025rzd}, the pair-antisymmetric case has a far more popular gravitational motivation in the context of extending general relativity (GR)~\cite{Einstein:1915}. Specifically, this index structure may be used to encode the spacetime \emph{torsion} tensor~$\MAGT{_{\mu}^\alpha_{\nu}}\equiv-\MAGT{_{\nu}^\alpha_{\mu}}$, whose geometric interpretation is illustrated in~\cref{NonRiemannianSchematic}. Recall that the curvature of spacetime is defined as
\begin{align}
\MAGF{_{\mu\nu}^\rho_\sigma} &\equiv 2\left(\tensor{\partial}{_{[\mu}}\MAGA{_{\nu]}^\rho_\sigma}+\MAGA{_{[\mu|}^\rho_\alpha}\MAGA{_{|\nu]}^\alpha_\sigma}\right) \,, \label{MAGFDef}
\end{align}
where~$\MAGA{_{\mu}^\rho_\nu}$ is the connection field, which is not a tensor owing to its inhomogeneous transformation under diffeomorphisms. In spacetimes where torsion is not assumed to vanish, it has the definition
\begin{equation}
	\MAGT{_\mu^\alpha_\nu} \equiv 2\MAGA{_{[\mu|}^\alpha_{|\nu]}} \,, \label{MAGTDef}
\end{equation}
i.e. it is a tensor that can be extracted out of the non-tensorial connection. This can be equivalently understood by partitioning~$\MAGA{_{\mu}^\rho_\nu}$ into the connection-valued Levi--Civita part~$\Christoffel{_\mu^\nu_\rho}\equiv\MAGg{^{\nu\lambda}}\big(\PD{_{(\mu}}\MAGg{_{\rho)\lambda}}-\frac{1}{2}\PD{_{\lambda}}\MAGg{_{\mu\rho}}\big)$, corresponding to the standard Christoffel symbols where~$\MAGg{_{\mu\nu}}$ is the metric tensor, plus a tensor-valued linear function of the torsion known as the \emph{contorsion} tensor
\begin{subequations}
\begin{align}
	\MAGA{_\mu^\rho_\nu} &\equiv \Christoffel{_\mu^\rho_\nu} +\Dis{_\mu^\rho_\nu}\,,\label{postriem}\\
	\Dis{_{\alpha\beta\gamma}} &\equiv \frac{1}{2}\left(\MAGT{_{\alpha\beta\gamma}} - \MAGT{_{\alpha\gamma\beta}} + \MAGT{_{\beta\alpha\gamma}}\right).\label{Contorsion}
\end{align}
\end{subequations}
It can be easily verified that~\cref{MAGTDef,postriem,Contorsion} are mutually consistent. The inclusion of torsion paints an attractive picture in~\cref{NonRiemannianSchematic} of a generalised, post-Einsteinian spacetime geometry. Despite its popular appeal, this narrative alone is not a firm ground upon which to engage in responsible model building~\cite{Barker:2025xzd}. The \emph{compelling} reason for considering torsion does not come from geometry, which already served out its usefulness as the classical inspiration for GR. Rather, torsion emerges from that perennial principle which knits both GR and the standard model of particle physics into the quantum framework of EFT: the principle of gauge theory. Specifically, torsion emerges as a covariant field-strength tensor in Poincar\'e gauge theory (PGT), as pioneered by Utiyama, Kibble and Sciama~\cite{Utiyama:1956sy,Kibble:1961ba,Sciama:1962}. 

\paragraph*{General Lagrangian} As shown in~\cite{Barker:2025xzd}, if torsion is propagating at all, and if it is perturbative, then its universal infrared (IR) limit is captured by the quadratic theory of a pair-antisymmetric rank-three field in flat space~$\MAGg{_{\mu\nu}}\to\G{_{\mu\nu}}$. In this limit, there is no reference to actual gravity whatsoever, which would otherwise appear as the dynamical perturbation of the metric~$\MAGg{_\mu_\nu}$ around the background $\G{_{\mu\nu}}$. Up to parity-violating operators, and physics living on the boundary, this universal IR theory (which we denote~\NamedModel{A23} so as to distinguish it from various special cases) has Lagrangian
\begin{align}
	\Lagrangian{A23} & \equiv
	\MAGG{2}{_1}\Lambda^2\Dis{_{\alpha\beta\chi}}\Dis{^{\alpha\beta\chi}}
	+\MAGG{2}{_2}\Lambda^2\Dis{_{\alpha\beta\chi}}\Dis{^{\beta\alpha\chi}}
	\nonumber\\& \hspace{5pt}
	+\MAGG{2}{_3}\Lambda^2\Dis{^{\alpha}_{\alpha}^{\beta}}\Dis{^{\chi}_{\beta}_{\chi}}
	+\MAGG{4}{_1}\PD{_{\beta}}\Dis{^{\delta}_{\chi}_{\delta}}
		\PD{^{\chi}}\Dis{^{\alpha}_{\alpha}^{\beta}}
	\nonumber\\& \hspace{5pt}
	+\MAGG{4}{_2}\PD{_{\chi}}\Dis{^{\delta}_{\beta}_{\delta}}
		\PD{^{\chi}}\Dis{^{\alpha}_{\alpha}^{\beta}}
	+\MAGG{4}{_3}\PD{_{\beta}}\Dis{^{\alpha\beta\chi}}
		\PD{_{\delta}}\Dis{_{\alpha\chi}^{\delta}}
	\nonumber\\& \hspace{5pt}
	+\MAGG{4}{_4}\PD{_{\alpha}}\Dis{^{\alpha\beta\chi}}
		\PD{_{\delta}}\Dis{_{\beta\chi}^{\delta}}
	+\MAGG{4}{_6}\PD{_{\beta}}\Dis{^{\alpha\beta\chi}}
		\PD{_{\delta}}\Dis{_{\chi\alpha}^{\delta}}
	\nonumber\\& \hspace{5pt}
	+\MAGG{4}{_7}\PD{^{\chi}}\Dis{^{\alpha}_{\alpha}^{\beta}}\PD{_{\delta}}\Dis{_{\chi\beta}^{\delta}}
	+\MAGG{4}{_9}\PD{_{\alpha}}\Dis{^{\alpha\beta\chi}}\PD{_{\delta}}\Dis{^{\delta}_{\beta\chi}}
	\nonumber\\& \hspace{5pt}
	+\MAGG{4}{_{15}}\PD{_{\delta}}\Dis{_{\alpha\beta\chi}}\PD{^{\delta}}\Dis{^{\alpha\beta\chi}}
	+\MAGG{4}{_{16}}\PD{_{\delta}}\Dis{_{\beta\alpha\chi}}\PD{^{\delta}}\Dis{^{\alpha\beta\chi}}
	\, ,\label{RootTheoryA23}
\end{align}
where the canonical dimension of~$\Dis{_{\alpha\beta\chi}}$ is necessarily unity.

\paragraph*{Previous attempts} We will dedicate part of~\cref{Sec:Concl} to performing a literature review of torsionful models that might connect to~\cref{RootTheoryA23} as non-linear completions, but before proceeding we particularly emphasise the linear analysis of Lin, Hobson and Lasenby in~\cite{Lin:2018awc,Lin:2019ugq} which, though not part of the present series, is very closely aligned. In that work the authors use spectral methods similar to those introduced in~\cref{Sec:Methods} to study a flat, quadratic model similar to that presented in~\cref{RootTheoryA23}. The important differences are twofold. Firstly, the model in~\cite{Lin:2018awc,Lin:2019ugq} includes not only the contorsion but also the metric perturbation,\footnote{More correctly, the fields used in~\cite{Lin:2018awc,Lin:2019ugq} are the spin connection and the tetrad perturbation, but the extra constraint of Poincar\'e gauge invariance makes this distinction physically irrelevant.} and also couplings between the two. Whereas we have discarded such couplings in~\cite{Barker:2025xzd} as being irrelevant in the IR, the use of both fields in~\cite{Lin:2018awc,Lin:2019ugq} stems from the assumption of a specialised non-linear completion of the quadratic model. This non-linear completion, to be introduced in~\cref{PGTAction}, actually \emph{restricts} the basis of kinetic operators to be smaller than that in~\cref{RootTheoryA23}. The specific choice of non-linear completion also accounts for the second difference between~\cite{Lin:2018awc,Lin:2019ugq} and the present letter: the catalogue in~\cite{Lin:2018awc,Lin:2019ugq} was obtained by tuning couplings to achieve simultaneous unitarity of the spectrum \emph{and} power-counting renormalisability, where the non-linear completion is absolutely needed to make claims about the latter. Generically, the linear models obtained in this way will feature symmetries which are not respected by the non-linear operators that were used to motivate them in the first place. The consequences of this are discussed above, and we aim to do better with the current approach. Not only do we require the models to be healthy already in their non-gravitating limit (a very basic consequence of the universal coupling to gravity), but our starting operator basis in~\cref{RootTheoryA23} is \emph{fully general}, and \emph{completely agnostic} about the non-linear completion. The non-linear completion, if any can be found at all, is expected to be meticulously dictated order-by-order through a consistent deformation process \cite{Barnich:1993vg}. This is a process which cannot be avoided, and which does not need any direction from geometric insight.

\paragraph*{In this letter} The remainder of this letter has the following structure. In~\cref{Sec:Methods} we recap the theory behind the algorithm used for systematically deriving all physically motivated special cases of~\cref{RootTheoryA23}. The exposition is brief, because this content was presented in full already in the prequel letter~\cite{Barker:2025rzd}. In~\cref{Sec:Concl}, the results of the survey are presented and their physical implications are discussed. The complete catalogue resulting from this work can be found in~\cref{Sec:Parameters}. We take~$\hbar\equiv c\equiv 1$, and the high-energy physics convention~$(+,-,-,-)$. Otherwise, we try to adhere to the conventions of~\cite{Percacci:2020ddy}.

\begin{figure*}[htbp]
	\includegraphics[height=\textheight]{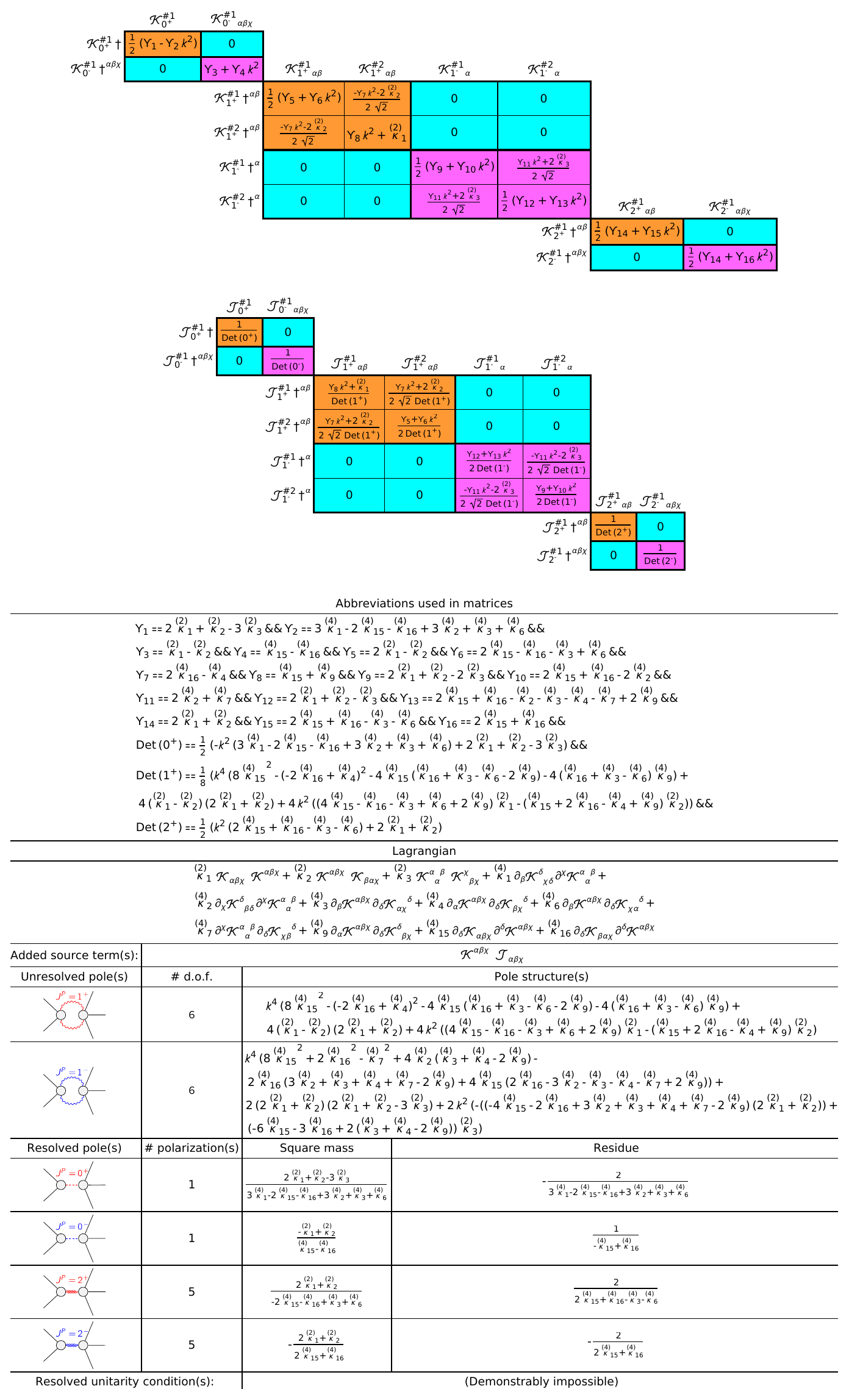}
	\caption{Output generated by \PSALTer{}. Spectrograph for the unconstrained and \emph{evidently non-unitary} theory~\NamedModel{A23} in~\cref{RootTheoryA23}. The upper matrices represent the wave operator blocks~$\mathsf{O}_{J^P}$ from~\cref{BlockDiagonal} --- the lower matrices show the pseudoinverses~$\mathsf{O}^+_{J^P}$. See~\cref{FieldKinematicsA23Field} for notational details; in all spectrographs the cutoff~$\Lambda^2$ is absorbed into the now-mass-dimension-two couplings~$\big\{\MAGG{2}{_i}\big\}$.}
\label{ParticleSpectrographA23}
\end{figure*}

\section{Theoretical development}\label{Sec:Methods}

\paragraph*{Wave operator} To evaluate the tree-level spectrum for each of the symmetric specializations of~\cref{RootTheoryA23}, we utilize the \textit{PSALTer} software package~\cite{Barker:2024juc,Barker:2025qmw}. The core of this method is the study of the propagator associated with a given Lagrangian~$\mathcal{L}_{n\leq4}$. We understand this to depend on the field d.o.f in~$\Dis{_{\alpha\beta\chi}}$, including derivatives, along with the parameters~$\big\{\MAGG{n}{_i}\big\}$, and~$\Lambda$. While it might seem unusual to parameterise the mass terms in the quadratic action as multiples of the cutoff scale~$\Lambda$, this nomenclature is justified in an EFT context, where the smallness of the~$\big\{\MAGG{2}{_i}\big\}$ plays an important perturbative role, as in the case of small quark masses in chiral perturbation theory~\cite{Gasser:1983yg}. We introduce, as test-fields, the sources~$\Cur{^{\alpha\beta\chi}}$ through the definition~$\mathcal{S}\equiv\int\mathrm{d}^4x\big[\mathcal{L}_{n\leq4}-\Dis{_{\alpha\beta\chi}}\Cur{^{\alpha\beta\chi}}\big]$. For the computational implementation, the independent d.o.f of~$\Dis{_{\alpha\beta\chi}}$ and the source~$\Cur{^{\alpha\beta\chi}}$ are represented by column vectors, denoted~$\mathsf{K}$ and~$\mathsf{J}$ respectively. In position space, this action can be written schematically as
\begin{equation}\label{GenLag}
	\mathcal{S}=\int\mathrm{d}^4x\ \mathsf{K}^{\text{T}}(x)\cdot\Big[\mathsf{O}(\partial)\cdot\mathsf{K}(x)-\mathsf{J}(x)\Big].
\end{equation}
The wave operator~$\mathsf{O}(\partial)$ is a polynomial tensor, at most quadratic in the derivative operator~$\PD{_\mu}$, whose coefficients depend on the couplings~$\big\{\MAGG{n}{_i}\big\}$ and are parametrised by the cutoff scale.

\paragraph*{Saturated propagator} The saturated propagator corresponding to the action in~\cref{GenLag} is determined in momentum space, where the Fourier transform is effected by the replacement~$\PD{_\mu}\mapsto -i\tensor{k}{_\mu}$. The calculation proceeds by inverting the wave operator and contracting it with the sources, as given in
\begin{equation}\label{Propagator}
	\Pi(k)\equiv\mathsf{J}^\dagger(k)\cdot\mathsf{O}^{-1}(k)\cdot\mathsf{J}(k).
\end{equation}
The special cases of~\cref{RootTheoryA23} we study are, by construction, endowed with additional gauge symmetries. A direct consequence of these symmetries is the degeneracy of the wave operator, which makes its formal inverse in~\cref{Propagator} singular. This issue is resolved because the same symmetries impose constraints on the source currents, ensuring that the divergences from the inverse are nullified in the `sandwiching' procedure of~\cref{Propagator}. We then apply standard polology principles to extract the physical content: poles in~$\Pi(k)$ correspond to particle masses --- which must be real to exclude tachyons --- and the associated residues must be positive definite to preclude ghosts.

\paragraph*{Massive spin and parity} For massive states, the analysis is greatly facilitated by employing a basis of definite spin and parity~($J^P$), which simplifies the wave operator inversion into a set of more manageable linear algebra problems. In this basis, the wave operator block-decomposes into a direct sum over these irreducible subspaces
\begin{equation}\label{BlockDiagonal}
	\mathsf{O}(k)\equiv\bigoplus_{J,P}\tensor{\mathsf{O}}{_{J^P}}(k).
\end{equation}
The spectroscopic analysis proceeds independently in each~$J^P$ subspace, which \PSALTer{} automatically identifies for the field~$\Dis{_{\alpha\beta\chi}}$, as detailed in~\cref{FieldKinematicsA23Field}. With this decomposition, we compute the spectrograph for the unconstrained theory~\NamedModel{A23} in~\cref{RootTheoryA23}, shown in~\cref{ParticleSpectrographA23}. All poles in the general theory are found to be massive. The~$1^+$ and~$1^-$ sectors each contain a pair of propagating states, whose squared masses are non-rational functions of the couplings~$\big\{\MAGG{n}{_i}\big\}$, as they derive from a quadratic equation in~$k\equiv\sqrt{\tensor{k}{^\nu}\tensor{k}{_\nu}}$. To avoid cumbersome symbolic manipulations, \PSALTer{} designates these poles as `unresolved'. The pathology of the theory is nevertheless evident from the `resolved' poles: the~$0^+$, $0^-$, $2^+$ and~$2^-$ sectors exhibit conflicting no-ghost and no-tachyon conditions for their respective single poles. This finding reinforces the argument from~\cref{Sec:Introduction} that gauge symmetries are essential. Imposing such symmetries will characteristically render some states massless, as in the familiar Fierz--Pauli and Proca--Maxwell examples, thereby requiring a subsequent analysis of massless poles.

\paragraph*{Massless helicity} The analysis of massless particles requires a shift from the spin-parity~($J^P$) basis, appropriate for massive states, to one of definite helicity~($h$). While the \PSALTer{} software computes the number and unitarity of massless modes via a component-wise analysis in the light-like frame~$\left[\tensor{k}{^\mu}\right] = \left(\omega, 0, 0, \omega\right)$, it does not intrinsically resolve their helicity content. Following~\cite{Barker:2025rzd} we extend the methodology of~\cite{Barker:2024juc,Barker:2025qmw} by constructing the helicity eigenstates from the source components. The procedure generalizes the well-known case of a vector source~$\Cur{^\mu}$, for which the helicity-$\pm1$ states are the combinations~$\hel{\pm 1}(k) \equiv \Cur{^1}(k) \pm i \Cur{^2}(k)$. For the rank-three tensor source~$\Cur{^{\alpha \beta \gamma}}(k)$, with the symmetry assumption following from~\cref{eq:DisA23} that~$\Cur{^{\alpha \beta \gamma}}(k)\equiv-\Cur{^{\alpha\gamma\beta}}(k)$, the corresponding decomposition follows from standard angular momentum rules, and this procedure results in the formulae
 \begin{subequations}
\begin{align} 
\hel[1]{\pm 2} &\equiv \Cur{^{223}} -\Cur{^{113}}\pm i (\Cur{^{123}} + \Cur{^{213}}), \label{Rank3HelFirst} \\
\hel[2]{\pm 2} &\equiv \Cur{^{202}}-\Cur{^{101}}\pm i (\Cur{^{102}} + \Cur{^{201}}), \\
\hel[1]{\pm 1} &\equiv \Cur{^{323}} \pm i\Cur{^{313}}, \\
\hel[2]{\pm 1} &\equiv \Cur{^{023}} \pm i\Cur{^{013}}, \\
\hel[3]{\pm 1} &\equiv \Cur{^{212}} \pm i\Cur{^{112}}, \\
\hel[4]{\pm 1} &\equiv \Cur{^{302}} \pm i\Cur{^{301}}, \\
\hel[5]{\pm 1} &\equiv \Cur{^{203}} \pm i\Cur{^{103}}, \\
\hel[6]{\pm 1} &\equiv \Cur{^{002}} \pm i\Cur{^{001}}, \\
\hel[1]{0} &\equiv \Cur{^{113}}+\Cur{^{223}}, \\
\hel[2]{0} &\equiv \Cur{^{123}}-\Cur{^{213}}, \\
\hel[3]{0} &\equiv \Cur{^{312}}, \\
\hel[4]{0} &\equiv \Cur{^{012}}, \\
\hel[5]{0} &\equiv \Cur{^{303}}, \\
\hel[6]{0} &\equiv \Cur{^{101}} + \Cur{^{202}}, \\
\hel[7]{0} &\equiv \Cur{^{201}} - \Cur{^{102}}, \\
\hel[8]{0} &\equiv \Cur{^{003}}. \label{Rank3HelLast}
\end{align}
\end{subequations}
Thus, for the 24 independent components of the pair-antisymmetric source~$\Cur{^{\alpha \beta \chi}}(k)$, the combinations with definite helicity~$h$ are denoted by~$\hel[d]{h}$, where the superscript~$d$ distinguishes between degenerate states. To apply these formulae to a specific model, one must account for the reduction in d.o.f imposed by gauge symmetries. The component-wise procedure in \PSALTer{} already manages these gauge constraints, which facilitates the construction of source combinations that are simultaneously gauge- and helicity-invariant. Converting~$\Pi(k)$ into this basis via the inverse of~\crefrange{Rank3HelFirst}{Rank3HelLast} allows for an unambiguous identification of the quantum numbers. This final step is not yet automated in \PSALTer{}, and for this reason the remaining spectrographs~\cref{ParticleSpectrographA23B1D1E1G2H1I1K2,ParticleSpectrographA23B1D1E1G2H1J1K1,ParticleSpectrographA23B1D1E1G2H1J2} for massless models shown in this letter do not resolve the helicity content explicitly. Nevertheless, we have performed the analysis manually to verify our claims in~\cref{Sec:Concl} regarding the spin sector~$J$ from which these modes originate.

\begin{figure*}[htbp]
\includegraphics[width=\linewidth]{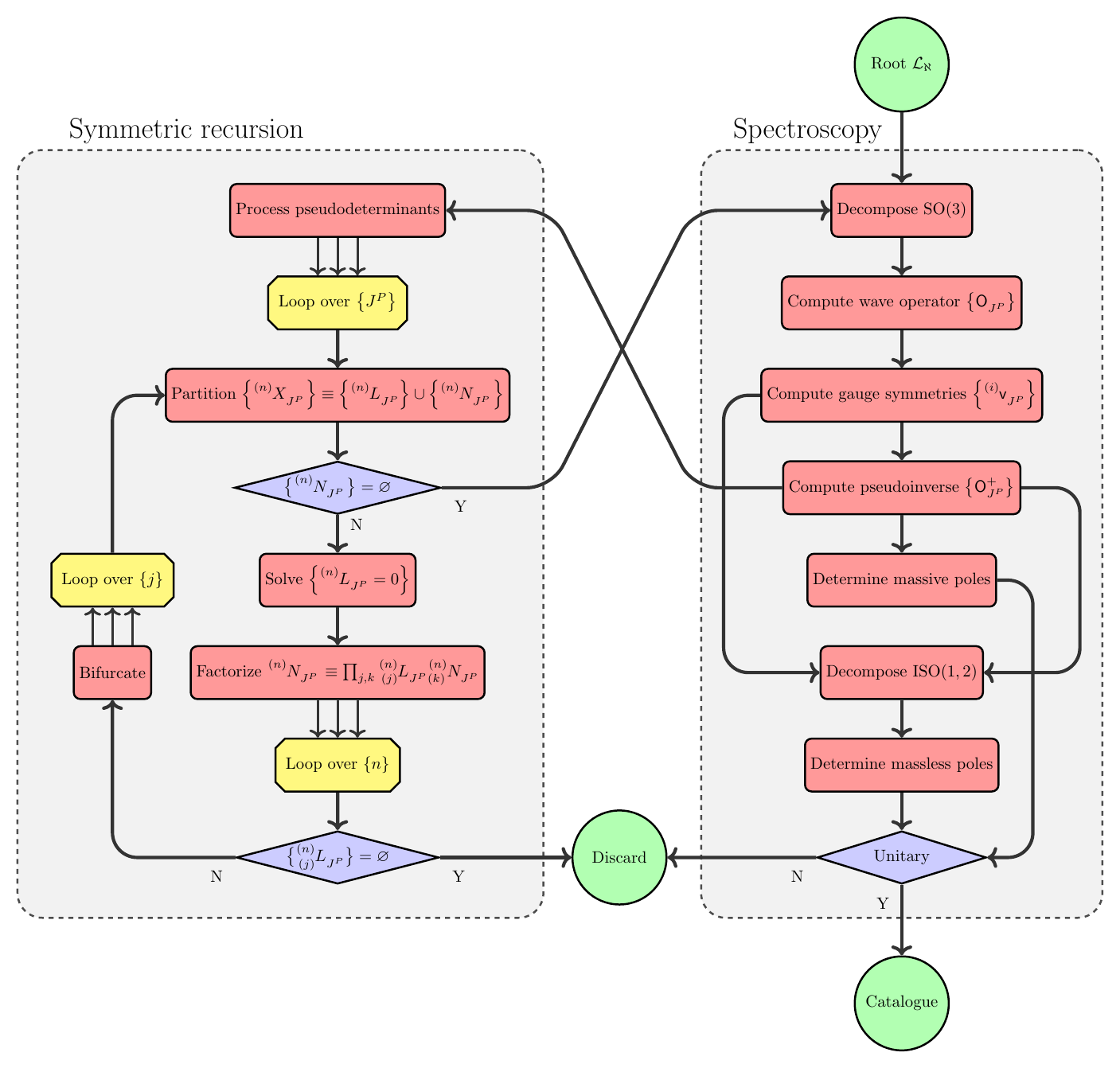}
	\caption{\label{Algorithm} The flowchart illustrates our method for systematically surveying all physically-motivated special cases of the general theory~$\mathcal{L}_{\aleph}$ in~\cref{RootTheoryA23}. The spectroscopy provides the spin-parity decomposed wave operator, which is used by the symmetry-search algorithm. The final output is an exhaustive catalogue of theories, as presented in~\cref{GraphRepresentationA23} and~\cref{AllModelsA23}.}
\end{figure*}

\paragraph*{Symmetries and constraints} To systematically identify all specialisations of the theory with enhanced gauge symmetry, we implement another different set of methodologies. This procedure begins with the momentum-space representation of the classical equations of motion, which derive from~\cref{GenLag}:
\begin{equation}\label{FieldEquations}
	\mathsf{O}(k)\cdot\mathsf{K}(k)=\frac{1}{2}\mathsf{J}(k).
\end{equation}
As can be seen in~\cref{FieldEquations}, a gauge invariance is manifested by the existence of principal null directions~$\GenNullVector{i}(k)$ which satisfy~$\mathsf{O}(k)\cdot\GenNullVector{i}(k)\equiv0$. These null vectors generate non-physical shifts in the field solution,~$\mathsf{K}(k) \mapsto \mathsf{K}(k) + \delta\mathsf{K}(k)$, where~$\delta\mathsf{K}\equiv\sum_ic_i(k)\GenNullVector{i}(k)$ and the~$c_i(x)$ are local gauge parameters. The existence of these null vectors implies that~$\mathsf{O}(k)$ is singular, obstructing the direct inversion required to compute the propagator in~\cref{Propagator}. As already discussed, this issue is circumvented because the symmetry also imposes constraints on the source current~$\mathsf{J}(k)$. Assuming a Hermitian wave operator, its right null eigenvectors are also left null eigenvectors. Left-multiplying the field equations in~\cref{FieldEquations} by~$\GenNullVector{i}^\dagger(k)$ thus leads to the constraints~$\GenNullVector{i}^\dagger(k)\cdot\mathsf{J}(k)=0$, which guarantee that the would-be divergences in the propagator cancel.

\paragraph*{Regularised propagator} Equipped with this understanding, the physical poles can then be extracted from the pseudodeterminant of a \emph{regularised} operator~$\Pi(k)\propto\det\tilde{\mathsf{O}}^{-1}(k)$ constructed using the invertible matrix
\begin{equation}\label{PseudoDeterminant}
	\tilde{\mathsf{O}}(k)\equiv\mathsf{O}(k)+\sum_i\GenNullVector{i}(k)\cdot\GenNullVector{i}^\dagger(k).
\end{equation}
The emergence of a new gauge symmetry is encoded in the vanishing of the pseudodeterminant,~$\det\tilde{\mathsf{O}}(k)$. This quantity can be expressed as a polynomial in momentum,~$\det\tilde{\mathsf{O}}(k)=\sum_n \GenXExpr{n}k^n$, where the coefficients~$\left\{\GenXExpr{n}\right\}$ are themselves polynomials in the Lagrangian couplings. A more specialized, symmetric model is therefore defined by any set of couplings that simultaneously solves the system of equations~$\left\{\GenXExpr{n}=0\right\}$.

\paragraph*{Symmetry bifurcation} These algebraic constraints are not guaranteed to be linear. When non-linear constraints appear, they may factorize, causing the solution space to bifurcate into distinct branches of theories. To systematically navigate this structure, our algorithm prioritizes any linear relationships among the couplings. This motivates a partition of the full set of constraints~$\left\{\GenXExpr{n}\right\}$ into linear and non-linear subsets
\begin{align}
	\left\{\GenXExpr{n}\right\} & \equiv \left\{\GenLExpr{n}\right\}\cup\left\{\GenNExpr{n}\right\}.
\end{align}
The search for symmetric specializations is implemented in practice by leveraging the spin-parity~($J^P$) decomposition of the wave operator, as shown in~\cref{BlockDiagonal}. The algorithm proceeds independently on each block~$\tensor{\mathsf{O}}{_{J^P}}(k)$, whose null vectors~$\big\{\NullVector{J}{P}{i}(k)\big\}$ produce the set of constraint equations~$\big\{\XExpr{J}{P}{n}=0\big\}$. These are then partitioned into their linear~$\big\{\LExpr{J}{P}{n}=0\big\}$ and non-linear~$\big\{\NExpr{J}{P}{n}=0\big\}$ components. The linear system is addressed first, and its unique solution is substituted into the non-linear constraints. This may cause algebraic factorization, revealing new linear conditions and thus branching points in the space of theories. This recursive process continues until no further linear factors can be extracted. Models defined solely by linear constraints are physically robust because they are independent of coupling redefinitions. Conversely, any irreducible non-linearities that remain must be resolved by an arbitrary choice of linear `slicings', introducing ambiguity. This ambiguity is not necessarily a physical pathology: it simply means that the gauge symmetry is realised by a parametric condition among the couplings. Since we are not aware of any precedent for such models in nature, we are unsure about their behaviour under renormalisation and simply quarantine them in our analysis. In~\cite{Barker:2025rzd} we found that, when applying this algorithm to the case of a totally symmetric rank-three field, the quarantined theories could be counted, and shown not to be predominant. For the theory in~\cref{RootTheoryA23}, however, the complexity of the branches is very much greater: we thus postpone even the counting of inherently non-linear or `parametric' models, to future work. The overall procedure is depicted in~\cref{Algorithm}.

\begin{figure*}[h]
	\AnnotatedGraph[%
		B = {0.187}{0.63}{1.02},%
		C = {0.265}{0.795}{1.02},%
		D = {0.34}{0.91}{1.02},%
		E = {0.420}{0.97}{1.02},%
		F = {0.5}{0.99}{1.02},%
		G = {0.577}{0.92}{1.02},%
		H = {0.655}{0.92}{1.02},%
		I = {0.735}{0.9}{1.02},%
		J = {0.812}{0.84}{1.02},%
		K = {0.89}{0.76}{1.02},%
	]{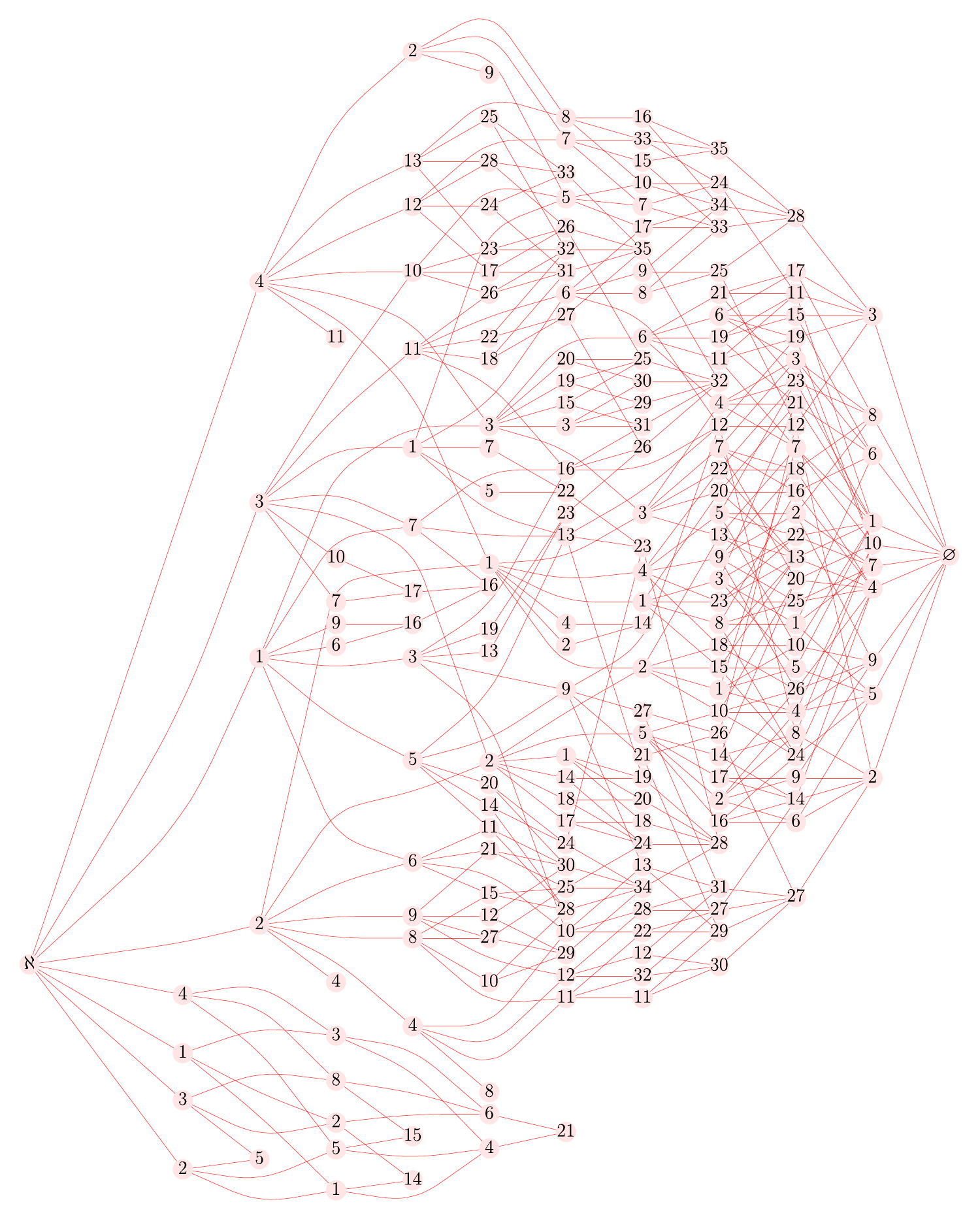}
	\caption{\label{GraphRepresentationA23} The complete classification of models for a pair-antisymmetric rank-three field. Each vertex corresponds to a separately defined theory, derived by imposing progressively more linear constraints (progressing from left to right) on the unconstrained theory~\NamedModel{A23} of~\cref{RootTheoryA23}. This process terminates at the theory \NamedModel{A23Z1}, for which~$\Lagrangian{A23Z1}\equiv 0$. The connections represent the constraints that define a daughter model in terms of its parent --- we show only connections spanning the fewest possible generations. A complete listing of all models is provided in~\cref{AllModelsA23}.}
\end{figure*}

\section{Results and discussion}\label{Sec:Concl}

\begin{figure*}[h]
	\includegraphics[width=\linewidth]{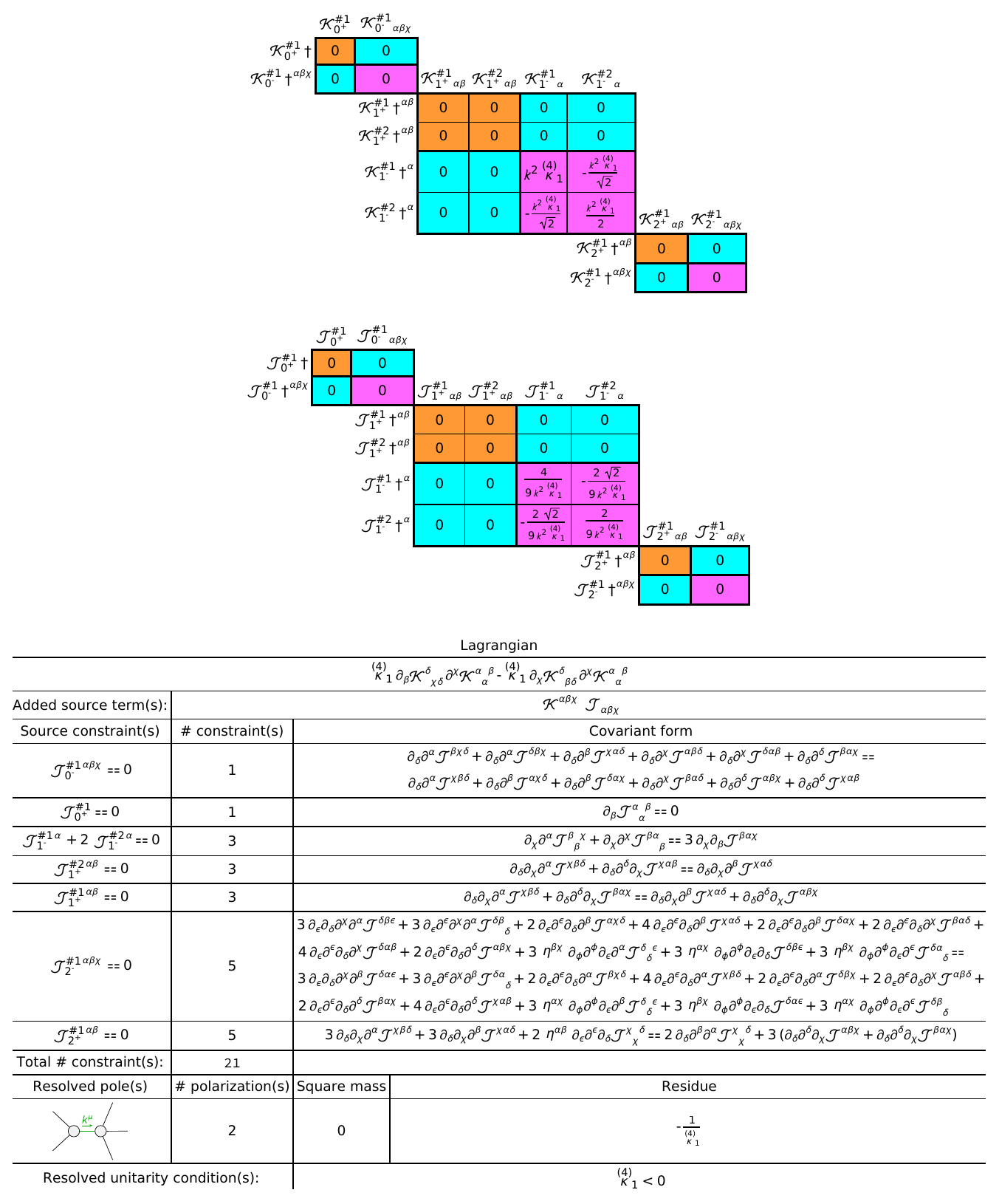}
	\caption{Output generated by \PSALTer{}. The spectrograph of~\NamedModel{A23B1D1E1G2H1I1K2}, as defined in~\cref{AllModelsA23}. All notation is defined in~\cref{FieldKinematicsA23Field,ParticleSpectrographA23}. This is an example of a model with minimal spin-one propagation, i.e. a single vector particle with one free parameter.}
\label{ParticleSpectrographA23B1D1E1G2H1I1K2}
\end{figure*}

\begin{figure*}[h]
	\includegraphics[width=\linewidth]{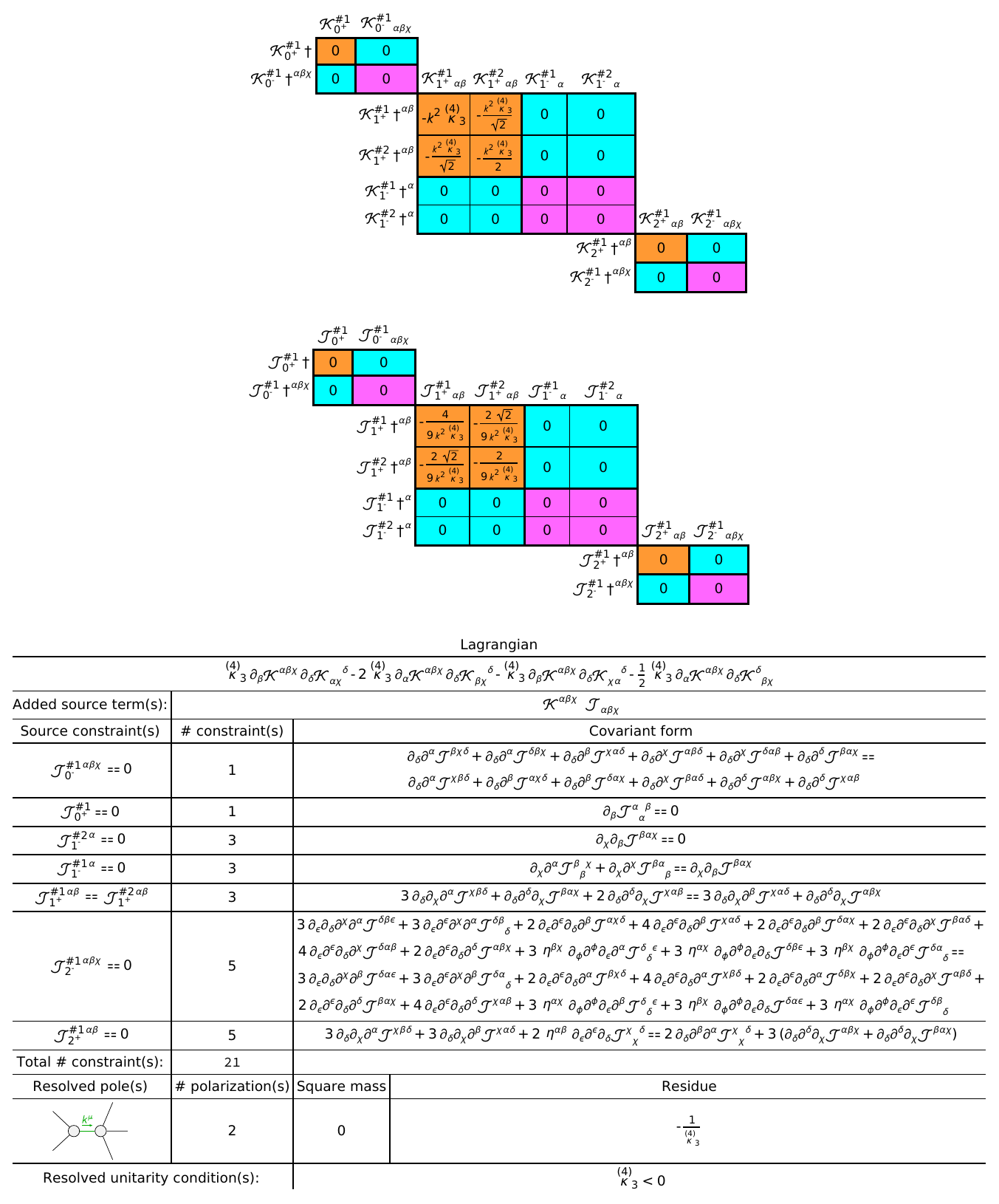}
	\caption{Output generated by \PSALTer{}. The spectrograph of~\NamedModel{A23B1D1E1G2H1J1K1}, as defined in~\cref{AllModelsA23}. All notation is defined in~\cref{FieldKinematicsA23Field,ParticleSpectrographA23}. As with~\NamedModel{A23B1D1E1G2H1I1K2} in~\cref{ParticleSpectrographA23B1D1E1G2H1J1K1}, this is another example of a model with minimal spin-one propagation, but this time the propagating vector is axial (even parity).}
\label{ParticleSpectrographA23B1D1E1G2H1J1K1}
\end{figure*}

\begin{figure*}[h]
	\includegraphics[width=\linewidth]{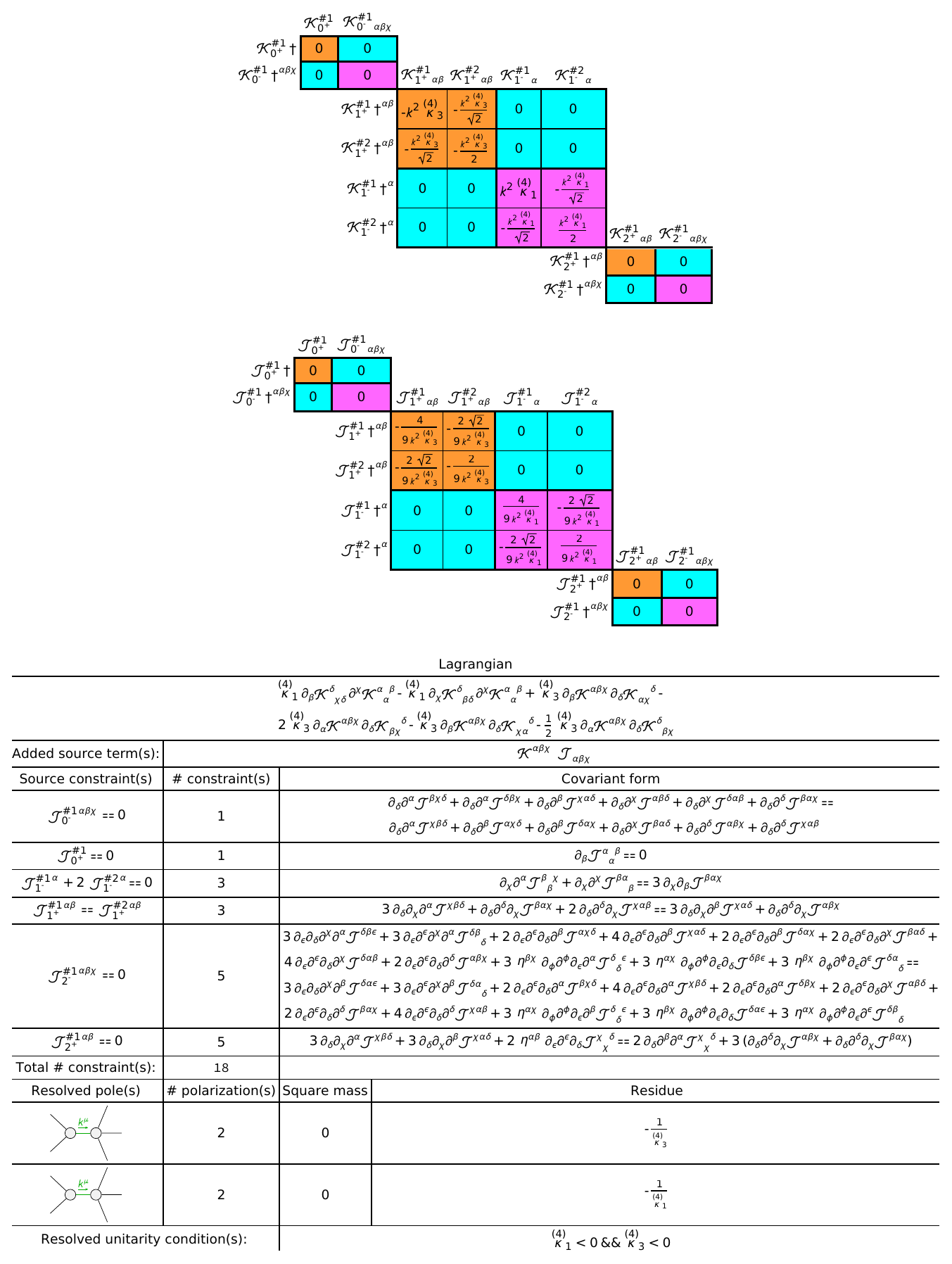}
	\caption{Output generated by \PSALTer{}. The spectrograph of~\NamedModel{A23B1D1E1G2H1J2}, as defined in~\cref{AllModelsA23}. All notation is defined in~\cref{FieldKinematicsA23Field,ParticleSpectrographA23}. This is an example of a model with minimal double spin-one propagation, i.e. a pair of vector particles with two free parameters, one for each sector (polar and axial).}
\label{ParticleSpectrographA23B1D1E1G2H1J2}
\end{figure*}

\paragraph*{Scale of catalogue} Based on the algorithm in~\cref{Sec:Methods} we extract 206 symmetric models whose degree of tuning is greater than for the root theory~\NamedModel{A23} and lesser than for the trivial empty theory~\NamedModel{A23Z1}. The graph of all models is illustrated in~\cref{GraphRepresentationA23}. We allocate a letter to indicate the number of constraints ($\mathfrak{B}$ means two constraints,~$\mathfrak{C}$ means three constraints, etc.) combined with a subscript to number different models. The full definition of each model is provided in~\cref{AllModelsA23}. Whilst our analysis of the actual particle spectra is not exhaustive,\footnote{We note that the small fraction of models for which unitarity conditions could not be assessed was excluded due to their highly complex structure, involving numerous parameters and multi-particle (more than just two) propagation. This complexity made assessing unitarity computationally impractical within reasonable time constraints.} as it was for the prequel letter~\cite{Barker:2025rzd}, we can fully profile the resulting \FinalTally{} unitary cases, which are:~\NamedModel{A23B1D1E1G1}, \NamedModel{A23B1D1E1G2H1I1K2}, \NamedModel{A23B1D1E1G2H1J1}, \NamedModel{A23B1D1E1G2H1J1K1}, \NamedModel{A23B1D1E1G2H1J2}, \NamedModel{A23B1D1E1G2I2}, \NamedModel{A23B1D1F2H1}, \NamedModel{A23B1D1F2H1I2}, \NamedModel{A23B1D1F2H2}, \NamedModel{A23B1D3F3H1I1}, \NamedModel{A23B1D3F3H2I2}, \NamedModel{A23B1D3F4H1I2}, \NamedModel{A23B1D4F1H2J2}, \NamedModel{A23B1E2G1H2}, \NamedModel{A23B2C1E2G2H1J2}, \NamedModel{A23B2C1E2G2I1}, \NamedModel{A23B3D2F2H1}, \NamedModel{A23B3D2F2H2}, \NamedModel{A23B4D4F3H2J1}, \NamedModel{A23B4E2G1H2}, \NamedModel{A23C2E1G2I2}, and~\NamedModel{A23C4E4G2I1}. As key examples, the complete spectrographs of~\NamedModel{A23B1D1E1G2H1I1K2}, \NamedModel{A23B1D1E1G2H1J1K1} and~\NamedModel{A23B1D1E1G2H1J2} are illustrated in~\cref{ParticleSpectrographA23B1D1E1G2H1I1K2,ParticleSpectrographA23B1D1E1G2H1J1K1,ParticleSpectrographA23B1D1E1G2H1J2}, respectively. To give our discussion of these results some context, we will focus mostly on the implications for theories with \emph{torsion}, including a brief literature review of such models. Despite this focus, it should nonetheless be borne in mind (see~\cref{Sec:Introduction}) that the IR limit in~\cref{RootTheoryA23} is relevant for any theory of a pair-antisymmetric rank-three field, of which torsion is but one example.

\paragraph*{Non-propagating torsion} Before discussing our results in detail, we recap the current status of torsion in the literature. This is important, as we imagine the IR limits derived from~\cref{RootTheoryA23} to be embedded in some non-linear completion in which gravity is playing an actual role. The baseline torsion model is the minimal Einstein--Cartan theory (ECT)~\cite{Cartan:1922,Cartan:1923,Cartan:1924,Cartan:1925,Einstein:1925, Einstein:1928,Einstein:19282}. ECT is the special case of PGT which `just happens' (in a sense whose compatibility with EFT principles has yet to be made clear) to share the Einstein--Hilbert action of GR
\begin{equation}\label{ECT}
	\mathcal{L}_{\text{ECT}}\equiv\frac{\MAGCouplingA{0}}{2}\sqrt{-g}\MAGF{},
\end{equation}
where~$\MAGF{}\equiv\MAGF{^{\mu\nu}_{\mu\nu}}$ is the scalar curvature, and we refer back to~\cref{MAGFDef} for the definition of the curvature tensor. Without coupling to matter,~\cref{ECT} simply implies the vacuum Einstein-like equations~$\MAGF{_{\lambda\mu}^\lambda_{\nu}}=0$ together with the auxiliary equation~$\MAGT{_{\mu}^\lambda_{\nu}}= 0$. Coupling to fermionic matter results in a contact spin-torsion interaction, whereby the torsion integrates out algebraically to leave effective four-Fermi interactions~\cite{Kibble:1961ba, Rodichev:1961,Freidel:2005sn,Alexandrov:2008iy,Shaposhnikov:2020frq,Karananas:2021zkl,Rigouzzo:2023sbb}. These effects have been put forward as having many phenomenological applications, for example in~\cite{Freidel:2005sn,Bauer:2008zj,Poplawski:2011xf,Diakonov:2011fs,Khriplovich:2012xg,Magueijo:2012ug,Khriplovich:2013tqa,Markkanen:2017tun,Carrilho:2018ffi,Enckell:2018hmo,Rasanen:2018fom,BeltranJimenez:2019hrm,Rubio:2019ypq,Shaposhnikov:2020geh,Karananas:2020qkp,Langvik:2020nrs,Shaposhnikov:2020gts,Mikura:2020qhc,Shaposhnikov:2020aen,Kubota:2020ehu,Enckell:2020lvn,Iosifidis:2021iuw,Bombacigno:2021bpk,Racioppi:2021ynx,Cheong:2021kyc,Dioguardi:2021fmr,Piani:2022gon,Dux:2022kuk,Rigouzzo:2022yan,Pradisi:2022nmh,Salvio:2022suk,Rasanen:2022ijc,Gialamas:2022xtt,Gialamas:2023emn,Gialamas:2023flv,Piani:2023aof,Poisson:2023tja,Rigouzzo:2023sbb,Barker:2023fem,Karananas:2023zgg,Martini:2023apm,He:2024wqv}. Despite these applications, and despite evidently being within the scope of~\cref{RootTheoryA23}, it is not immediately clear how non-propagating torsion can be reconciled conceptually with the EFT framework in~\cref{EFTLag}.\footnote{Given the large literature claiming to extract predictive results from non-propagating torsion, the fact that any such model is unlikely to deviate from the pre-existing EFT of metric-based gravity coupled to matter, once the torsion has been integrated out, seems concerning. This question will be investigated in further work.} Among a large literature, one particular body of work stands out for its fairly recent attempts at making this reconciliation. This avenue of work was opened up with the observation that the cutoff scale of Higgs inflation becomes elevated when the connection is taken to be independent of the metric; in essence, an equivalent way of positing~\cref{postriem} known historically as the \emph{Palatini} formulation~\cite{Bauer:2010jg,Shaposhnikov:2020geh,Shaposhnikov:2020fdv}. The basic Palatini Higgs model extends~\cref{ECT} by a non-minimal coupling between the bilinear of the Higgs doublet and~$\MAGF{}$. It was proposed in~\cite{Karananas:2021zkl} to extend this model by including all other operators with the lowest mass dimension which do not endow~$\MAGT{_{\mu}^\lambda_{\nu}}$ with a propagator pole.\footnote{This process was extended beyond torsion to fully non-metric metric-affine geometries in~\cite{Rigouzzo:2022yan,Rigouzzo:2023sbb}.} This approach is evidently close in spirit to ours, but since we exclude from our final tally of unitary models the `boring' parts of the catalogue with completely non-propagating torsion (trivial, empty spectra according to~\PSALTer{}), we do not connect with it.

\paragraph*{Propagating torsion} By contrast with~\cref{ECT}, propagating torsion is usually motivated by the action
\begin{align}\label{PGTAction}
	&\mathcal{L}_{\text{PGT}}\equiv-\frac{1}{2}\sqrt{-g}\Big[
		-\MAGCouplingA{0}\MAGF{}
		+\MAGF{^{\mu\nu\rho\sigma}}\big(
		\MAGCouplingG{1}\MAGF{_{\mu\nu\rho\sigma}}
		\nonumber\\& \hspace{5pt}
		+\MAGCouplingG{3}\MAGF{_{\rho\sigma\mu\nu}}
		+\MAGCouplingG{4}\MAGF{_{\mu\rho\nu\sigma}}
		\big)
		+\MAGFFirstThird{^{\mu\nu}}\big(
		\MAGCouplingG{7}\MAGFFirstThird{_{\mu\nu}}
		\nonumber\\& \hspace{5pt}
		+\MAGCouplingG{8}\MAGFFirstThird{_{\nu\mu}}
		\big)
		+\MAGCouplingG{16}\MAGFFirstThird{}^2
		+\MAGT{^{\mu\rho\nu}}\big(
		\MAGCouplingB{1}\MAGT{_{\mu\rho\nu}}
		\nonumber\\& \hspace{5pt}
		+\MAGCouplingB{2}\MAGT{_{\mu\nu\rho}}
		\Big)
	+\MAGCouplingB{3}\MAGT{^\mu}\MAGT{_\mu}
	\Big],
\end{align}
where two new contractions are defined as~$\MAGFFirstThird{_{\mu\nu}}\equiv\MAGF{_{\lambda\mu}^\lambda_\nu}$ and~$\MAGT{_\mu}\equiv\MAGT{_\lambda^\lambda_\mu}$.\footnote{In~\cref{ECT,PGTAction} we take care to adhere to the conventions of the canonical reference~\cite{Percacci:2020ddy}. In particular~$\MAGCouplingA{0}$ and the~$\left\{\MAGCouplingB{i}\right\}$ are of mass dimension two, whilst the~$\left\{\MAGCouplingG{i}\right\}$ are dimensionless.} The logic underpinning the non-linear completion in~\cref{PGTAction} is that one should restrict to a Yang--Mills-type action which is up to quadratic order in the field-strength tensors concomitant with gauging the Poincar\'e group~\cite{Hayashi:1967se,Hayashi:1980qp,Yo:1999ex,Yo:2001sy,Puetzfeld:2004yg}. Harkening back to~\cref{Sec:Introduction}, this is precisely the non-linear completion which is assumed in~\cite{Lin:2018awc,Lin:2019ugq}. Among the~$\MAGCouplingA{0}$,~$\left\{\MAGCouplingB{i}\right\}$ and~$\left\{\MAGCouplingG{i}\right\}$ are only ten couplings, short of the 12 couplings in~\cref{RootTheoryA23}. A vast literature is devoted to arbitrary tunings of these couplings so as to produce ghost-tachyon-free linear spectra~\cite{Sezgin:1981xs,Blagojevic:1983zz,Blagojevic:1986dm,Kuhfuss:1986rb,Yo:1999ex,Yo:2001sy,Blagojevic:2002,Puetzfeld:2004yg,Yo:2006qs,Shie:2008ms,Nair:2008yh,Nikiforova:2009qr,Chen:2009at,Ni:2009fg,Baekler:2010fr,Ho:2011qn,Ho:2011xf,Ong:2013qja,Puetzfeld:2014sja,Karananas:2014pxa,Ni:2015poa,Ho:2015ulu,Karananas:2016ltn,Obukhov:2017pxa,Blagojevic:2017ssv,Blagojevic:2018dpz,Tseng:2018feo,Lin:2018awc,BeltranJimenez:2019acz,Zhang:2019mhd,Aoki:2019rvi,Zhang:2019xek,Jimenez:2019qjc,Lin:2019ugq,Percacci:2019hxn,Barker:2020gcp,BeltranJimenez:2020sqf,MaldonadoTorralba:2020mbh,Barker:2021oez,Marzo:2021esg,Marzo:2021iok,delaCruzDombriz:2021nrg,Baldazzi:2021kaf,Annala:2022gtl,Mikura:2023ruz,Mikura:2024mji,Barker:2024ydb,Karananas:2024xja}. We have argued in~\cref{Sec:Introduction} and~\cite{Barker:2025xzd,Barker:2025rzd} that this process is conceptually flawed, but the non-linear completion in~\cref{PGTAction} anyway leads to more problems even when the linear spectra are ghost- and tachyon-free. These non-linear problems manifest as strongly coupled particles~\cite{Moller:1961,Pellegrini:1963,Hayashi:1967se,Cho:1975dh, Hayashi:1979qx,Hayashi:1979qx,Dimakis:1989az,Dimakis:1989ba,Lemke:1990su,Hecht:1990wn,Hecht:1991jh,Yo:2001sy,Afshordi:2006ad,Magueijo:2008sx,Charmousis:2008ce,Charmousis:2009tc,Papazoglou:2009fj,Baumann:2011dt,Baumann:2011dt,DAmico:2011eto,Gumrukcuoglu:2012aa,Wang:2017brl,Mazuet:2017rgq,BeltranJimenez:2020lee,JimenezCano:2021rlu,Barker:2022kdk,Delhom:2022vae,Annala:2022gtl,Barker:2022kdk,Barker:2023fem,Karananas:2024hoh}, whose presence suggests that the afflicted theory is inherently non-perturbative near the Minkowski background upon which it was first tuned~\cite{Vainshtein:1972sx,Deffayet:2001uk,Deffayet:2005ys,Charmousis:2008ce,Charmousis:2009tc,Papazoglou:2009fj,deRham:2014zqa,Deser:2014hga,Wang:2017brl}. Any gauge symmetries which do emerge through the arbitrary tuning process are also typically broken at non-linear order (so-called `accidental' symmetries)~\cite{Velo:1969txo,Aragone:1971kh,Cheng:1988zg,Hecht:1996np,Chen:1998ad,Yo:1999ex,Yo:2001sy,Blixt:2018znp,Blixt:2019ene,Blixt:2020ekl,Krasnov:2021zen,Bahamonde:2021gfp,Delhom:2022vae}.\footnote{See also~\cite{Hayashi:1980qp,Blagojevic:1983zz,Blagojevic:1986dm,Yo:2001sy,Blagojevic:2002,Ong:2013qja,Blagojevic:2013dea,Blagojevic:2013taa,Blagojevic:2018dpz,BeltranJimenez:2019hrm,Aoki:2020rae,Barker:2021oez}.} Since the seminal work of Yo and Nester a quarter of a century prior~\cite{Yo:1999ex,Yo:2001sy}, the consensus (see e.g.~\cite{Hecht:1996np,Chen:1998ad,Yo:1999ex,Yo:2001sy}) has been that:\footnote{See also~\cite{Yo:2006qs,Shie:2008ms,Chen:2009at,Baekler:2010fr,Ho:2011qn,Ho:2011xf,Ho:2015ulu,Tseng:2018feo,Zhang:2019mhd,Zhang:2019xek,MaldonadoTorralba:2020mbh,delaCruzDombriz:2021nrg,Karananas:2025xcv,Shaposhnikov:2025znm} for various applications,~\cite{Puetzfeld:2004yg,Ni:2009fg,Puetzfeld:2014sja,Ni:2015poa,Barker:2020gcp} to review the literature, and~\cite{BeltranJimenez:2019acz,BeltranJimenez:2019esp,Percacci:2020ddy,BeltranJimenez:2020sqf,Marzo:2021esg,Piva:2021nyj,Marzo:2021iok,Iosifidis:2021xdx,Jimenez-Cano:2022sds,Iosifidis:2023pvz} for analyses of spacetime non-metricity which are more or less analogous.}
\begin{quote}
	\emph{The only consistent Einstein--Cartan theories or Poincar\'e gauge theories are those in which the spin-zero scalar or pseudoscalar torsion is allowed to propagate.}
\end{quote}
The spin-zero parts of torsion correspond to the scalar~$0^+$ and pseudoscalar~$0^-$ modes in~\cref{FieldKinematicsA23Field}, which together span the first diagonal blocks of the wave operator and propagator in~\cref{ParticleSpectrographA23}. Indeed, by examining~\cref{ParticleSpectrographA23}, we see that in the general IR foundation these modes both try to propagate, respectively with masses
\begin{subequations}
\begin{align}
	m_{0^+}^2 &\equiv \frac{\Lambda^2\left(2\MAGG{2}{_1}+\MAGG{2}{_2}-3\MAGG{2}{_3}\right)}{3\MAGG{4}{_1}-2\MAGG{4}{_{15}}-\MAGG{4}{_{16}}+3\MAGG{4}{_2}+\MAGG{4}{_3}+\MAGG{4}{_6}},\label{Mass0p} \\
	m_{0^-}^2 &\equiv \frac{\Lambda^2\left(\MAGG{2}{_2}-\MAGG{2}{_1}\right)}{\MAGG{4}{_{15}}-\MAGG{4}{_{16}}}.\label{Mass0m}
\end{align}
\end{subequations}
Notwithstanding these poles, the underlying theory of~\cref{RootTheoryA23} is as a whole non-unitary. We have challenged the paradigm of (pseudo)scalar-only torsion at the classical level with two recent proposals for \emph{vector} torsion originating in the~$1^+$ and~$1^-$ sectors of~\cref{FieldKinematicsA23Field}.\footnote{The proposals~\cite{Barker:2023fem,Barker:2024goa} do not form part of the present series.} Interestingly, these vector modes, which populate the second diagonal blocks in~\cref{ParticleSpectrographA23}, clearly account for the richest sector of the IR limit of torsion in~\cref{RootTheoryA23}: both parities are able to simultaneously propagate pairs of particles with distinct masses given by the solutions to a quadratic equation in~$k^2$. Our previous proposals for sourcing particles from this sector were as follows:
\begin{description}
	\item[More fields] In~\cite{Barker:2023fem} it was shown using a fully non-perturbative Hamiltonian analysis that strongly-coupled torsion modes could be pacified by extra multiplier fields, leading to an Einstein--Proca theory from vector torsion.
	\item[Bigger gauge group] In~\cite{Barker:2024goa} it was shown that the Maxwell-like operator for~$\MAGT{_\mu}$ --- which is missing from~\cref{PGTAction} but reached by setting~$\MAGG{4}{_1}+\MAGG{4}{_2}=0$ in~\cref{RootTheoryA23} --- could be motivated as a Yang--Mills-type term when the Poincar\'e gauge group is enlarged to the full conformal group.
\end{description}
Both of these proposals implicitly `buy' the idea that the restriction to the Yang--Mills-type action in~\cref{PGTAction} is physically well motivated, rather than being assumed by some loose analogy. Indeed, this idea has itself been challenged by the authors of~\cite{Percacci:2020ddy} and in recent efforts~\cite{Baldazzi:2021kaf,Martini:2023apm,Melichev:2023lwj,Melichev:2025hcg} in which gravitational model-building has been forced to be reconciled with the most basic expectations of quantum field theory. The objective of the present series is to fall in line with this predictive (albeit arduous) framework, and we will now see that it provides the strongest motivation for propagating \emph{vector} torsion to date.

\paragraph*{Science products} The systematic identification of unitary and symmetric models within the pair-antisymmetric tensor field~$\Dis{_{\alpha\beta\gamma}}$ yields multiple independent configurations exhibiting significantly greater complexity compared to the previously studied and well-known case of the totally symmetric tensor. However, such complexity is not reflected in the type of propagation, which invariably involves vector particles (of either parity). It is a critical outcome of our study that within the sub-set of \FinalTally{} models for which the unitarity is assessed:\footnote{Hopes of recovering simple, unitary single-particle propagation from the complex models where unitarity could not be assessed are naturally low, given the observed presence of multiple poles across many different sectors.}
\begin{quote}
	\emph{We do not find any symmetry which can support the tuning necessary for propagating spin-zero or spin-two torsion.}
\end{quote}
Indeed, no known examples of symmetric propagation exhibiting these features exist in the literature. Instead, the complex nature of the \FinalTally{} identified unitary models is reflected in various distinctive and novel frameworks supporting vector propagation. Starting from the simplest scenario, spin-one propagation with only one free parameter --- analogous to familiar implementations of lower rank --- the models extend to more sophisticated configurations with multiple parameters (and thus independent combinations of operators), resulting, in some cases, in the simultaneous propagation of a pair of spin-one modes. The ultimate measure of a model's computational usefulness depends on its ability to support consistent interactions and to yield non-trivial theories. This likely requires dedicated studies for each of the distinct IR foundations identified. Here, we provide an initial broad classification of the \FinalTally{} models, categorised by their propagation characteristics and overall complexity, the latter simply assessed by counting the number of free coefficients involved. We emphasise that this is a preliminary classification with no sharp boundaries.

\paragraph*{Minimal spin one} We find four models that minimally propagate a vector particle. A glance at the spectrograph in~\cref{ParticleSpectrographA23B1D1E1G2H1I1K2} reveals that~\NamedModel{A23B1D1E1G2H1I1K2} is the simplest implementation of~$1^-$ propagation. The same propagation is afforded by model~\NamedModel{A23B2C1E2G2H1J2}, which adds a decoupled and non-propagating scalar sector, allowed by a less constraining gauge invariance. An axial vector is instead propagated by model~\NamedModel{A23B1D1E1G2H1J1K1} --- see~\cref{ParticleSpectrographA23B1D1E1G2H1J1K1} --- and, with a less constraining symmetry, by~\NamedModel{A23B1D1E1G2H1J1}. 

\paragraph*{Non-minimal spin one} In the presence of multiple parameters, a thorough study might help in revealing possible reductions by highlighting the irrelevance and decoupling of particular combinations. Here we resort to a simpler measure, based on how many of the surviving Lagrangian parameters appear in the unitarity conditions. This simplified approach selects~\NamedModel{A23B1D1E1G1} as a particularly involved model which, nevertheless, is symmetric, unitary, and propagates only one (vector) sector. While further study might reveal some simplification, the unitarity constraints involve all five parameters of the corresponding Lagrangian. Slightly less intricate is model~\NamedModel{A23B3D2F2H2}, with four parameters. This category is further populated by the remaining set~\NamedModel{A23B1D1E1G2I2}, \NamedModel{A23B1D1F2H1}, \NamedModel{A23B1D1F2H1I2}, \NamedModel{A23B1D3F3H1I1}, \NamedModel{A23B1D3F3H2I2}, \NamedModel{A23B1D3F4H1I2}, \NamedModel{A23B1D4F1H2J2}, \NamedModel{A23B1E2G1H2}, \NamedModel{A23B2C1E2G2I1}, \NamedModel{A23B4D4F3H2J1}, \NamedModel{A23B4E2G1H2}, \NamedModel{A23C2E1G2I2} and~\NamedModel{A23C4E4G2I1}. All these models involve up to three parameters and non-trivial unitarity conditions. We reiterate that, for the simplest of these cases, an overlap with the previous category can occur.

\paragraph*{Minimal double spin one} Multiple simultaneous propagation is arguably one of the more interesting mechanisms unleashed by the use of high-rank fields. While often competing, with conflicting requirements of unitarity, it is known that different sectors might live in mutual harmony already for the totally symmetric case~\cite{Francia:2013sca,Barker:2025rzd}. In the larger space of a pair-antisymmetric field, we find three apparently different models which simultaneously propagate two vectors. As illustrated in~\cref{ParticleSpectrographA23B1D1E1G2H1J2}, the model~\NamedModel{A23B1D1E1G2H1J2} is particularly simple, with just two parameters, one for each sector (polar and axial), without any intertwining of unitarity conditions. 

\paragraph*{Non-minimal double spin one} At the other end of the spectrum, models~\NamedModel{A23B1D1F2H2} and~\NamedModel{A23B3D2F2H1} have up to four parameters, all of which are involved in the complicated unitarity conditions. Part of the apparent growth in complexity for the last two models can be blamed on the reduced symmetry in the spin-two sector which, unlike for~\NamedModel{A23B1D1E1G2H1J2}, is not completely degenerate. 

\paragraph*{Further work} Whilst~\cref{GraphRepresentationA23} and~\cref{AllModelsA23} should exhaustively catalogue the IR foundations associated with the field defined in~\cref{eq:DisA23}, the increased complexity of the system leads to limitations in our subsequent analysis compared to that performed in~\cite{Barker:2025rzd}, namely:
\begin{description}
	\item[Unitarity] It is possible that our final tally of \FinalTally{} unitary models out of 206 candidate IR foundations in~\cref{GraphRepresentationA23,AllModelsA23} may be slightly increased in the future. This is because a small minority of models in the catalogue resisted automatic unitarity analysis by \PSALTer{}.\footnote{Somewhat confusingly, we count 22 such treatment-resistant models: precisely the same number as are confirmed to be unitary! To clarify why the analysis fails in these cases: the final step in the \PSALTer{} algorithm is to reduce the gathered-together system of inequality constraints for positive pole residues and real masses. This final step relies on certain proprietary functionality in \Wolfram{}, here running on a single AMD Ryzen 7 6800U CPU core, and is set to timeout after an arbitrary period of 20 seconds. We do not suggest that these models are in any way exceptional physically.} Note that the number of new models, if any, is expected to be small: most of the viable IR foundations for parity-preserving torsion have been established in this work. 
	\item[Parametric models] We do not provide details of symmetric models which may be reached by the imposition of inherently non-linear constraints on the couplings; a concept introduced in~\cref{Sec:Methods}. Whether parametric models offer a robust IR foundation for constructing EFTs is a matter to be investigated separately, ideally by performing the renormalisation of a toy model.
\end{description}
Beyond these aspects of further work, the same extensions apply as were mentioned in~\cite{Barker:2025rzd}, namely the inclusion of parity-violating operators and the systematic non-linear completion of each model in our catalogue. The former task is straightforward, and has been facilitated by~\cite{Barker:2025qmw}. The latter task is highly non-trivial, and may well prove to be impossible for most of the catalogue. As a final comment, we reiterate that the critical need for reconciling torsionful model-building with the principles of quantum field theory has been highlighted in at least three branches of the literature, preliminarily in~\cite{Lin:2018awc,Lin:2019ugq}, and more thoroughly in~\cite{Baldazzi:2021kaf,Martini:2023apm,Melichev:2023lwj,Melichev:2025hcg}, and separately in~\cite{Bauer:2010jg,Shaposhnikov:2020geh,Shaposhnikov:2020fdv,Karananas:2021zkl,Rigouzzo:2022yan,Rigouzzo:2023sbb} and related works (see most recently~\cite{Karananas:2025ews} and a particularly cogent review in~\cite{Shaposhnikov:2025znm}). It is to be hoped for a future integration between approaches, whose conclusions do not always align. Outside of this literature, it is helpful to remember that -- when engaging in responsible gravitational model-building -- quantum considerations are in no sense `optional'.

\begin{acknowledgments}
This work used the DiRAC Data Intensive service~(CSD3 \href{www.csd3.cam.ac.uk}{www.csd3.cam.ac.uk}) at the University of Cambridge, managed by the University of Cambridge University Information Services on behalf of the STFC DiRAC HPC Facility~(\href{www.dirac.ac.uk}{www.dirac.ac.uk}). The DiRAC component of CSD3 at Cambridge was funded by BEIS, UKRI and STFC capital funding and STFC operations grants. DiRAC is part of the UKRI Digital Research Infrastructure.

This work also used the Newton compute server, access to which was provisioned by Will Handley using an ERC grant.

W.~B. is grateful for the support of Girton College, Cambridge, Marie Skłodowska-Curie Actions and the Institute of Physics of the Czech Academy of Sciences. The work of C.~M. was supported by the Estonian Research Council grant PRG1677. A.~S. acknowledges financial support from the ANID CONICYT-PFCHA/DoctoradoNacional/2020-21201387.

Co-funded by the European Union (Physics for Future -- Grant Agreement No. 101081515). Views and opinions expressed are however those of the author(s) only and do not necessarily reflect those of the European Union or European Research Executive Agency. Neither the European Union nor the granting authority can be held responsible for them.
\end{acknowledgments}

\bibliography{Manuscript}

\appendix

\section{Catalogue}\label{Sec:Parameters}

This appendix presents, in~\cref{AllModelsA23}, the complete definitions for all models illustrated in the hierarchy of~\cref{GraphRepresentationA23}.

\def\arraystretch{1.5}
\begin{longtable*}{ @{\extracolsep{0pt}} >{\centering\arraybackslash}m{0.05\linewidth} | >{\centering\arraybackslash}m{0.475\linewidth} | >{\centering\arraybackslash}m{0.475\linewidth} }
\caption{This table provides a complete catalogue of all models depicted in the hierarchy of~\cref{GraphRepresentationA23}. For each entry, the second column specifies the defining constraints as vanishing linear combinations of the parameters in~\cref{RootTheoryA23}. The third column outlines the additional conditions that result in more specialized, symmetric theories. For the consistent theories~\NamedModel{A23B1D1E1G2H1I1K2}, \NamedModel{A23B1D1E1G2H1J1K1} and~\NamedModel{A23B1D1E1G2H1J2}, the corresponding spectrographs are provided in~\cref{ParticleSpectrographA23B1D1E1G2H1I1K2,ParticleSpectrographA23B1D1E1G2H1J1K1,ParticleSpectrographA23B1D1E1G2H1J2}.}\label{AllModelsA23}\\
\hline\hline
\# & Definition & Deconfliction \\
\hline
\endfirsthead
\multicolumn{3}{c}{Table \thetable{} -- continued from previous page} \\
\hline\hline
\endhead
\hline\hline
\multicolumn{3}{r}{Continued on next page}
\endfoot
\hline\hline
\endlastfoot
\NamedModel{A23} &~--- &~$\Big[ \MAGG{4}{_{15}} - \MAGG{4}{_{16}}=0 \wedge \MAGG{2}{_{1}} - \MAGG{2}{_{2}}=0 \Big] \vee \Big[ 2\MAGG{4}{_{15}} + \MAGG{4}{_{16}}=0 \wedge 2\MAGG{2}{_{1}} + \MAGG{2}{_{2}}=0 \Big] \vee \Big[ 2\MAGG{4}{_{15}} + \MAGG{4}{_{16}} - \MAGG{4}{_{3}} - \MAGG{4}{_{6}}=0 \wedge 2\MAGG{2}{_{1}} + \MAGG{2}{_{2}}=0 \Big] \vee \Big[ 3\MAGG{4}{_{1}} - 2\MAGG{4}{_{15}} - \MAGG{4}{_{16}} + 3\MAGG{4}{_{2}} + \MAGG{4}{_{3}} + \MAGG{4}{_{6}}=0 \wedge 2\MAGG{2}{_{1}} + \MAGG{2}{_{2}} - 3\MAGG{2}{_{3}}=0 \Big] \vee \Big[ 2\MAGG{4}{_{3}} + \MAGG{4}{_{4}} - 2\MAGG{4}{_{6}} + 4\MAGG{4}{_{9}}=0 \wedge -4\MAGG{4}{_{15}} - 2\MAGG{4}{_{16}} + \MAGG{4}{_{4}} - 4\MAGG{4}{_{9}}=0 \wedge 2\MAGG{2}{_{1}} + \MAGG{2}{_{2}}=0 \Big] \vee \Big[ 2\MAGG{4}{_{3}} + \MAGG{4}{_{4}} - 2\MAGG{4}{_{6}} + 4\MAGG{4}{_{9}}=0 \wedge \MAGG{4}{_{15}} - \MAGG{4}{_{16}} + \tfrac{1}{2}\MAGG{4}{_{4}} + \MAGG{4}{_{9}}=0 \wedge \MAGG{2}{_{1}} - \MAGG{2}{_{2}}=0 \Big] \vee \Big[ \MAGG{4}{_{3}} + \MAGG{4}{_{4}} + \tfrac{3}{2}\MAGG{4}{_{7}} - 2\MAGG{4}{_{9}}=0 \wedge 2\MAGG{4}{_{15}} + \MAGG{4}{_{16}} + \MAGG{4}{_{7}}=0 \wedge 2\MAGG{2}{_{1}} + \MAGG{2}{_{2}}=0 \Big] \vee \Big[ \MAGG{4}{_{3}} + \MAGG{4}{_{4}} + \tfrac{3}{2}\MAGG{4}{_{7}} - 2\MAGG{4}{_{9}}=0 \wedge 4\MAGG{4}{_{15}} + 2\MAGG{4}{_{16}} - 6\MAGG{4}{_{2}} - \MAGG{4}{_{7}}=0 \wedge 2\MAGG{2}{_{1}} + \MAGG{2}{_{2}} - 3\MAGG{2}{_{3}}=0 \Big]$ \\
\hline
\NamedModel{A23B1} &~$-\MAGG{4}{_{15}} + \MAGG{4}{_{16}}=0 \wedge -\MAGG{2}{_{1}} + \MAGG{2}{_{2}}=0$ &~$\Big[ \MAGG{4}{_{15}}=0 \wedge \MAGG{2}{_{1}}=0 \Big] \vee \Big[ 3\MAGG{4}{_{15}} - \MAGG{4}{_{3}} - \MAGG{4}{_{6}}=0 \wedge \MAGG{2}{_{1}}=0 \Big] \vee \Big[ 3\MAGG{4}{_{1}} - 3\MAGG{4}{_{15}} + 3\MAGG{4}{_{2}} + \MAGG{4}{_{3}} + \MAGG{4}{_{6}}=0 \wedge \MAGG{2}{_{1}} - \MAGG{2}{_{3}}=0 \Big]$ \\
\end{longtable*}

\end{document}